\PassOptionsToPackage{inline}{enumitem}
\PassOptionsToPackage{hyphens}{url}
\PassOptionsToPackage{x11names,dvipsnames}{xcolor}

\documentclass[twocolumn]{fairmeta}

\usepackage{xcolor}
\usepackage{graphicx}
\usepackage{transparent}
\usepackage{tikz}
\usepackage{babel}
\usepackage[normalem]{ulem}
\usepackage{xurl}
\usepackage{verbatim}
\usepackage{hyperref}
\usepackage{listings}
\usepackage{varwidth}
\usepackage[super]{nth}
\usepackage{moreenum}
\usepackage{tikz}
\usepackage{minted}
\usepackage[absolute,overlay]{textpos}
\usepackage{tabfigures}
\usepackage{pgfplots}
\usepackage{siunitx}
\usepackage{textcomp}
\usepackage{multirow}
\usepackage{pgfcalendar}
\usepackage{tabularx}
\usepackage{etoolbox}

\pgfplotsset{compat=1.18}

\usepgfplotslibrary{units}
\usepgfplotslibrary{dateplot}
\usepgfplotslibrary{colormaps}
\usepgfplotslibrary{groupplots}

\pgfplotsset{colormap/viridis high res}

\pgfplotsset{
    /pgfplots/layers/standard with background/.define layer set={
        background,
        axis background,
        axis grid,
        axis ticks,
        axis lines,
        axis tick labels,
        pre main,
        main,
        axis descriptions,
        axis foreground
    }{/pgfplots/layers/standard}
}

\pgfmathdeclarefunction{lg10}{1}{%
    \pgfmathparse{ln(#1)/ln(10)}%
}

\usetikzlibrary{
  decorations.pathreplacing,
  positioning,
  shapes,
  calc,
  fit,
  tikzmark,
  backgrounds,
  arrows.meta,
  intersections
}

\tikzset{
  white background/.style={
    show background rectangle,
    tight background,
    background rectangle/.style={
      fill=white
    }
  }
}

\tikzstyle{global}=[
  white background,
]

\tikzstyle{txt}=[align=center, execute at end node=\vphantom{bg}]
\tikzstyle{tiny}=[node font=\tiny]
\tikzstyle{ss}=[node font=\scriptsize]
\tikzstyle{fn}=[node font=\footnotesize]
\tikzstyle{sm}=[node font=\small]
\tikzstyle{tiny txt}=[tiny, txt]
\tikzstyle{fn txt}=[fn, txt]
\tikzstyle{mas}=[midway, above, sloped]
\tikzstyle{mbs}=[midway, below, sloped]

\colorlet{good}{green!75!black}
\colorlet{bad}{red!75!black}
\colorlet{ok}{yellow!75!black}

\tikzstyle{good}=[green!75!black]
\tikzstyle{bad}=[red!75!black]
\tikzstyle{ok}=[yellow!75!black]

\tikzset{>={Classical TikZ Rightarrow[]}}
\tikzset{small >/.tip={Classical TikZ Rightarrow[scale=0.67]}}
\tikzset{smaller >/.tip={Classical TikZ Rightarrow[scale=0.5]}}

\definecolor{fbblue}{rgb}{.278, .404, .667}

\colorlet{samplebg}{gray!50}
\colorlet{sample1}{fbblue}
\colorlet{sample2}{gray}
\colorlet{sample3}{green!45!white!60!black}
\colorlet{sample4}{red!45!white!65!black}

\colorlet{graymain}{gray}
\colorlet{graylight}{graymain!70!white}
\colorlet{graylighter}{graymain!45!white}
\colorlet{graylightest}{graymain!25!white}
\colorlet{graydark}{graymain!70!black}
\colorlet{graydarker}{graymain!45!black}

\colorlet{bluemain}{fbblue}
\colorlet{bluelight}{bluemain!70!white}
\colorlet{bluelighter}{bluemain!45!white}
\colorlet{bluelightest}{bluemain!25!white}
\colorlet{bluedark}{bluemain!70!black}
\colorlet{bluedarker}{bluemain!45!black}

\colorlet{greenmain}{green!45!white!60!black}
\colorlet{greenlight}{greenmain!70!white}
\colorlet{greenlighter}{greenmain!45!white}
\colorlet{greenlightest}{greenmain!25!white}
\colorlet{greendark}{greenmain!70!black}
\colorlet{greendarker}{greenmain!45!black}

\colorlet{redmain}{red!45!white!65!black}
\colorlet{redlight}{redmain!70!white}
\colorlet{redlighter}{redmain!45!white}
\colorlet{redlightest}{redmain!25!white}
\colorlet{reddark}{redmain!70!black}
\colorlet{reddarker}{redmain!45!black}

\colorlet{yellowmain}{yellow!80!white!75!black}
\colorlet{yellowlight}{yellowmain!70!white}
\colorlet{yellowlighter}{yellowmain!45!white}
\colorlet{yellowlightest}{yellowmain!25!white}
\colorlet{yellowdark}{yellowmain!70!black}
\colorlet{yellowdarker}{yellowmain!45!black}

\colorlet{problemred}{red!15!white}
\colorlet{cautionorange}{orange!25!white}
\colorlet{warningyellow}{yellow!50!white}
\colorlet{highlightgreen}{green!15!white}
\colorlet{highlightblue}{fbblue!15!white}

\setminted{style=sas}

\newminted{cpp}{fontsize=\small,linenos,numbersep=3pt,xleftmargin=12pt,autogobble,escapeinside=||,beameroverlays,frame=leftline,rulecolor=graylightest}

\newminted{python}{fontsize=\small,linenos,numbersep=3pt,xleftmargin=12pt,autogobble,escapeinside=||,beameroverlays,frame=leftline,rulecolor=graylightest}

\newminted{thrift}{fontsize=\small,linenos,numbersep=3pt,xleftmargin=12pt,autogobble,escapeinside=||,beameroverlays,frame=leftline,rulecolor=graylightest}

\newminted{json}{fontsize=\small,linenos,numbersep=3pt,xleftmargin=12pt,autogobble,escapeinside=||,beameroverlays,frame=leftline,rulecolor=graylightest}

\newminted{text}{fontsize=\small,linenos,numbersep=3pt,xleftmargin=12pt,autogobble,escapeinside=||,beameroverlays,frame=leftline,rulecolor=graylightest}

\newlist{todolist}{itemize}{2}
\setlist[todolist]{label=$\square$,leftmargin=0.7cm}
\usepackage{pifont}

\newcommand{\fakeinterrobang}{{\ooalign{?\cr\kern.2ex!\cr}}}

\newcommand{\PreserveBackslash}[1]{\let\temp=\\#1\let\\=\temp}
\newcolumntype{C}[1]{>{\PreserveBackslash\centering}p{#1}}
\newcolumntype{R}[1]{>{\PreserveBackslash\raggedleft}p{#1}}
\newcolumntype{L}[1]{>{\PreserveBackslash\raggedright}p{#1}}

\setenumerate[0]{label={\tbfigures\arabic*.}}

\def\NAME{OpenZL}

\usepackage{amsfonts}
\usepackage{algorithm}
\usepackage{algpseudocode}
\usepackage{nicefrac}

\usepackage{amsthm}
\theoremstyle{definition}
\newtheorem{definition}{Definition}[section]

\newtheorem*{remark}{Remark}

\usepackage[T1]{fontenc}
\usepackage[scale=0.95]{sourcecodepro}

\newlist{enuminline}{enumerate*}{1}
\setlist[enuminline]{label=(\roman*)}

\colorlet{openzlplotcolor}{ForestGreen}
\colorlet{zstdplotcolor}{Cyan}
\colorlet{xzplotcolor}{Fuchsia}
\colorlet{zlibplotcolor}{Mahogany}
\colorlet{brotliplotcolor}{Dandelion}
\colorlet{bloscplotcolor}{Magenta}
\colorlet{parquetplotcolor}{RoyalBlue}

\pgfdeclareplotmark{openzlplotmark}{\pgfuseplotmark{*}}
\pgfdeclareplotmark{zstdplotmark}{\pgfuseplotmark{triangle*}}
\pgfdeclareplotmark{xzplotmark}{\pgfuseplotmark{diamond*}}
\pgfdeclareplotmark{zlibplotmark}{\pgfuseplotmark{star}}
\pgfdeclareplotmark{brotliplotmark}{\pgfuseplotmark{pentagon*}}
\pgfdeclareplotmark{bloscplotmark}{\pgfuseplotmark{+}}
\pgfdeclareplotmark{parquetplotmark}{\pgfuseplotmark{x}}

\newif\ifcommentsenabled

\commentsenabledtrue

\newcommand\coloruline[1]{\bgroup\markoverwith{\textcolor{#1}{\rule[-0.5ex]{2pt}{1pt}}}\ULon}

\title{\NAME{}: A Graph-Based Model for Compression}

\author[]{Yann Collet}
\author[]{Nick Terrell}
\author[]{W.\ Felix Handte}
\author[]{Danielle Rozenblit}
\author[\S]{Victor Zhang}
\author[]{Kevin Zhang}
\author[]{Yaelle Goldschlag}
\author[]{Jennifer Lee}
\author[]{Elliot Gorokhovsky}
\author[]{Yonatan Komornik}
\author[*]{Daniel Riegel}
\author[*]{Stan Angelov}
\author[*]{Nadav Rotem}

\affiliation[]{Meta Platforms, Inc.}
\contribution[\S]{Manuscript author}
\contribution[*]{Joint last author}

\abstract{
Research techniques in the last decade have improved lossless compression ratios by significantly increasing processing time. These techniques have remained obscure because production systems require high throughput and low resource utilization. In practice, application-specific compression algorithms that leverage knowledge of the data structure and semantics are more popular. Application-specific compressor systems outperform even the best generic compressors, but these techniques have some drawbacks. Application-specific compressors are inherently limited in applicability, have high development costs, and are difficult to maintain and deploy.

In this work, we show that these challenges can be overcome with a new compression strategy. We propose the ``graph model'' of compression, a new theoretical framework for representing compression as a directed acyclic graph of modular codecs. OpenZL compresses data into a self-describing wire format, any configuration of which can be decompressed by a universal decoder. OpenZL’s design enables rapid development of tailored compressors with minimal code; its universal decoder eliminates deployment lag; and its investment in a well-vetted standard component library minimizes security risks. Experimental results demonstrate that OpenZL achieves superior compression ratios and speeds compared to state-of-the-art general-purpose compressors on a variety of real-world datasets. Internal deployments at Meta have also shown consistent improvements in size and/or speed, with development timelines reduced from months to days. OpenZL thus represents a significant advance in practical, scalable, and maintainable data compression for modern data-intensive applications.

}

\date{October \nth{6}, 2025}
\correspondence{Yann Collet at \email{cyan@meta.com}}
\metadata[Code]{\url{https://github.com/facebook/openzl}}
\metadata[Blog Post]{\url{https://engineering.fb.com/2025/10/06/developer-tools/openzl-open-source-format-aware-compression-framework}}

\begin{document}

\maketitle

\section{Introduction}
\label{sections:introduction}

Compression research over the last decade has largely focused on leveraging machine learning to improve compression ratios~\cite{KNOLL12, CMIX, NNCP, GOYAL20, LSTM-COMPRESS, MAO22}. This benefits scenarios where minimizing data size is critical but speed is less important~\cite{LIU22, LSTM-COMPRESS, NNCP, ZHANG24}. Neural-net-based approaches compress benchmark datasets at significantly better ratios than traditional techniques but achieve throughput on the order of 1KB/s and often require heavy GPU resources~\cite{LIU22, GOYAL20, CMIX, ZHANG24}.

By contrast, production workloads must strike a balance between compressed size and processing time. For this reason, almost all popular compressors on the market use a variant of LZ77~\cite{GUPTA17}, as its fast speeds and reasonable compression ratio makes it suitable for latency-sensitive, high-volume environments. Indeed, the last great leap in production-scale compression was Zstandard (Zstd)~\cite{zstandard}, which combines LZ77 with entropy coding, and whose typical usage compresses on the order of 100 MB/s and decompresses on the order of 1 GB/s. Many papers investigating compression for real-world applications say the quiet part out loud: neural-net-based systems are too slow and require too many resources to be serious contenders in production, despite superior ratios~\cite{JUMAR18, THEVENIN22, IQBAL20, 2BrightSparks24, LinuxReviews}.

As a consequence, domain-specific compressors are increasingly the tool of choice for data-intensive workflows. In fields like genomics~\cite{PIBIRI22, PIBIRI23, ALANKO25, CHANDAK18, LAN21, CHANDRA12}, computer graphics~\cite{MENTZER20, TAUBMAN19, Pranckevicius25, MESHOPTIMIZER}, and AI models~\cite{HERSHCOVITCH24}, tailored compression algorithms have pushed the state of the art in both academic and industry applications.

Across disparate read/write patterns, data lifetime requirements, and data organization, there is a clear through-line: knowing \emph{anything at all} about one's data yields better and faster compression than even the best generic compressors. And furthermore, the more structure that can be exploited out of the data, the better one can be on both axes.

This then begs the question: why aren't application-specific compressors more common? In other words, if there's a clear way to be both smaller and faster than generic compressors, \emph{why is it so rare?}

\begin{enumerate}[label=(\roman*)]
    \item \textbf{Upfront investment is intractable.} Designing and implementing compression algorithms requires significant expertise and effort. Zstd, for instance, contains over 100,000 lines of code, representing years of development and hand-optimization from a well-funded team. In addition, application-specific compression algorithms requires expertise not only in compression techniques, but also in the problem domain.
    \item\label{bad2} \textbf{Inflexibility of solutions.} Most custom compressors are designed to optimize for one benchmark dataset. Scale issues are immediate once you try to onboard datasets that require compression techniques not exploited by the original dataset. Additionally, just supporting new wire formats of the same data often requires extensive refactoring or even a complete rebuild of the library.
    \item \textbf{Security guarantees are hard to make.} A codebase as complex as a compression library is difficult to debug. Coupled with the fact that decompression is often performed on untrusted user input, this makes it very likely to contain intrusion channels that can be exploited.
\end{enumerate}

These constraints prevent smaller teams from adopting custom compression solutions. For larger enterprises that can fund their way out of these challenges, a different set of constraints arise when custom solutions are deployed at scale.

\begin{enumerate}[resume*]
    \item \textbf{Iteration is difficult.} Releasing a production library means freezing the wire format and publishing long-term support guidance. Oftentimes this inhibits the speed of new development, as backwards compatibility limits what you can ship. This reduces the efficacy of custom compression because not all gains can be realized.
    \item \textbf{Deployment is slow.} Updating from version $n$ to $n + 1$ requires that all the data readers are rolled out to support version $n+1$ before any of the writers are allowed to write the new version. Thus, without guarantees on library freshness, updating the wire format becomes untenable. This makes custom compression unsuitable for whole swaths of applications, including mobile app development and IoT devices.
    \item \textbf{High-cardinality applications are hard to support.} This is an extension of point \ref{bad2}, mostly focused on data warehouses with diverse customer needs. Deploying custom compressors for each of your clients' needs quickly becomes untenable once you scale past a handful of use cases.
\end{enumerate}

In this paper, we show that these challenges can be overcome with a new way of thinking about compression. We introduce \NAME{}, a general compression engine that uses a graph-structured compression model with a self-describing wire format and a universal decoder. Specifically, \NAME{} breaks from the typical monolithic compressor architecture by decomposing a compression into a DAG of composable \textit{codecs}.

\subsection{The Graph Model of Compression}

Our main theoretical contribution is the graph model of compression, defined formally in \cref{sections:concepts}. In summary, we define a \textit{compression graph} as a computational graph~\cite{COLLINS18} where the nodes are \textit{codecs} and edges represent data generated as output of one codec and used as input for another. \textit{Codecs} are defined simply to be multivariate functions, so are allowed to have multiple inputs and multiple outputs. The graph model allows us to think about compressors at a higher level of abstraction and enables new ways of approaching the task of compression.

The computational graph also means decompression is purely procedural. Apart from the compressed data itself, all you need is the graph to decode any valid compressed frame. The universality of the decoder is an important result of the graph model.

\subsection{Overview of Results}

This paper's main result is demonstrating that the graph model (as implemented) is both easy to use and expressive enough to cover a wide range of applications. \NAME{} is able to address several pain points of application-specific compressors in order to simplify their development:

\begin{enumerate}[label=(\roman*)]
    \item \textbf{Upfront investment is minimal.} Compared to existing monolithic compressor architectures, \NAME{}'s modular structure solves the flexibility problem and allows the same developer to quickly stand up compressors for many different formats. We show in \cref{sections:results} that the production code written to support new datasets is on the order of hundreds, and sometimes tens of lines of code.
    \item \textbf{Solutions are flexible.} The composable graph model also means diverse datasets can be supported by simply creating new graphs. Our benchmark experiments in \cref{sections:results} demonstrate that many disparate data formats can be easily parsed to take advantage of \NAME{}.
    \item \textbf{The security surface is minimized.} By breaking a compression job into small, independent chunks, we simplify the job of securing the entire compressor to simply securing each individual codec. \Cref{sections:components} outlines the security hardened \textbf{Standard Component Library} provided by the \NAME{} library. Compressors composed of these standard components inherit the strong security guarantees of the library.
\end{enumerate}

\NAME{} is also an enterprise-scale solution for custom compression.

\begin{enumerate}[resume*]
    \item \textbf{Iteration is easy.} The self-describing wire format means you can evolve a compressor's graph over time without needing to make any changes to the decoder.
    \item \textbf{Deployment is fast.} A universal decoder elides the rollout calculation. The same core library can decode any graph given to it.
    \item \textbf{High-cardinality applications are easy to support.} Serialized graphs are often on the order of kilobytes and can be deployed widely. Training can be exploited to scale compression ratio wins across many disparate datasets. \Cref{sections:openzl_at_meta} summarizes the adoption of \NAME{} within Meta and the performance wins we have enjoyed over time.
\end{enumerate}

\subsection{Paper Organization}

\Cref{sections:related_work} describes prior work on data compression, in particular composable compression engines.

\Cref{sections:concepts} develops the graph model of compression, the underlying theoretical framework for \NAME{}.

\Cref{sections:building_a_compressor} explores some common steps in building and training \NAME{} compressors.

\Cref{sections:implementation} explores the implementation and design choices of the \NAME{} code.

\Cref{sections:results} contains our experimental results, which show compressors built in \NAME{} that achieve ratio and speed improvements on a variety of different data.

\Cref{sections:openzl_at_meta} summarizes the benefits we've seen from using \NAME{} internally at Meta.

\Cref{sections:conclusions} contains some concluding remarks and a look to the future.

\section{Related Work}
\label{sections:related_work}

It is well known that a universal compressor that compresses every input does not exist. If some inputs are shortened, others are necessarily lengthened. Within those constraints, practical designs converge on a common recipe:
\begin{enuminline}
\item\label{traditional-stages-transform} apply reversible transforms to surface structure,
\item\label{traditional-stages-model} model the transformed symbols to estimate probabilities, and
\item\label{traditional-stages-encode} entropy-code them near optimally
\end{enuminline}.

A more interesting bound is the Shannon limit~\cite{SHANNON}, a measure of uncertainty that gives the maximum possible compression ratio for a given source distribution. In practice, this limit only matters for the last \ref{traditional-stages-encode} entropy-coding stage and doesn't tell you what the source is.

Significant effort goes into the modeling stage which reduces input into a residual error signal, for the following entropy stage to encode. An effective model minimizes the error and result in data that compresses well. 

The model does not necessarily need to operate directly on the input data. Various preparation stages can be inserted beforehand, as long as they are \ref{traditional-stages-transform} reversible transforms. Such transforms don’t necessarily shrink the data (though some do); they mostly massage it to make modelling more efficient.

Most compression systems follow this 3-stage design. This taxonomy structures our review. We briefly list representative transforms, point to modeling--coding pairs, note classic general-purpose LZ and prediction-centric families, and quickly cover systems that treat compression as a programmable composition of stages, foreshadowing the design targeted by \NAME{}.

\subsection{Reversible Transforms}

Reversible transforms are useful to remove redundancy and expose structure for downstream modeling. 

For example, the \textbf{Burrows--Wheeler Transform (BWT)}~\cite{burrows1994bwt} clusters similar contexts so nearby symbols look alike. The \textbf{run-length encoding (RLE)}~\cite{RLE} collapses symbol runs. The \textbf{LZ77} family~\cite{lz77} replaces repeated substrings with backward references. Match selection is non-trivial---overlaps and ties exist---and these choices materially affect downstream compressibility (e.g., shorter offsets, fewer distinct symbols). Another common strategy is \textbf{dictionary substitution}, where frequent substrings are factored into a table and the input is rewritten as indices into that table. Other examples include \textbf{delta coding} (replacing absolute values by differences) and \textbf{move-to-front (MTF) coding}~\cite{MTF}, which reorders symbols adaptively to expose locality for entropy coding.

None of these transforms have to reduce size on their own, even if some do. The point is rather to set up the following model stage for success, so that it can produce a more efficient residual signal.

\subsection{Entropy Coding and Modeling}

\paragraph{Modeling.}
The model~\cite{BLELLOCH13} turns the transformed stream into probabilities. Concretely, it defines the \emph{symbolization} exposed by upstream transforms (e.g., literals vs. match tuples), the \emph{conditioning} (contexts, orders, etc.), and the \emph{update law} (static per-block histograms, per-chunk adaptive counts, etc.).
It also budgets \emph{side information} (tables/parameters) so the bits spent on the model are paid back in the residual.

\paragraph{Entropy coding.}
Entropy coders map symbols to bitstreams given a probability distribution. \textbf{Huffman} remains widely used~\cite{huff}; \textbf{arithmetic coding} approaches the entropy limit with different latency/state trade-offs~\cite{WITTEN87}. The \textbf{ANS} family offers arithmetic-like compression at higher speed~\cite{duda2014}, with Zstandard’s FSE~\cite{FSE} as a table-driven tANS example. In practice, coders are mature; current work centers on high-throughput, optimized implementations and careful quantization of probabilities to the coder’s precision.

In short, the model does most of the work, provided that
\begin{enuminline}
\item upstream transforms expose the right symbols,
\item its probability estimates are expressed in the limited precision that the coder can handle (using smoothing or quantization so they stay well-behaved), and
\item side information (tables/params) is small enough and updated sensibly for the target speed and latency
\end{enuminline}.

\subsection{General-Purpose LZ}

For scenarios with strict throughput requirements, \textbf{LZ4}~\cite{lz4} and \textbf{Snappy}~\cite{snappy} apply LZ-style parsing with a lightweight tagged format (varint lengths/distances) rather than a full entropy-coding stage. They are effective for structured and textual data where modest ratios suffice but (de-)compression speed is critical.

\textbf{DEFLATE/gzip}~\cite{rfc1951, rfc1952} encodes literal/length and distance symbols, intertwined in a single Huffman-coded stream, using (static or dynamic) Huffman trees. \textbf{Brotli}~\cite{rfc7932} improves on gzip by using context models for literals and offsets. \textbf{Zstandard}~\cite{rfc8878} factors tokens into four logical streams (literals, literal-lengths, match-lengths, offsets) and uses either Huffman or FSE depending on mode. Zstandard further exploits parallelism at multiple levels (blocks, threads, and instruction-level). \textbf{LZMA} (as used in \textbf{xz}) combines LZ parsing and a range coder with context models (it is not a ``pure LZ'' design, making it more powerful but also markedly slower)~\cite{LZMA}.

\subsection{Prediction-Centric Compressors}

Higher compression ratios hinge on accurate symbol prediction. \textbf{PPM} conditions on preceding contexts~\cite{ppm}; \textbf{DMC} learns an adaptive automaton~\cite{dmc}; \textbf{context mixing} (e.g., \textbf{PAQ}~\cite{mahoney2013paq}, \textbf{cmix}~\cite{CMIX}) blends multiple predictors using ideas related to boosting. Today these are increasingly neural-net based. \textbf{NNCP} is a more recent approach in this vein~\cite{NNCP}.

Despite excellent ratios, these algorithms remain orders of magnitude slower (KB/s scale) and sequential, making them unsuitable for datacenter hot paths, where LZ compression techniques achieve orders of magnitude higher throughput.

\subsection{A Programmable Composition of Processing Stages}

Modern ``LZ + coder'' compressors already operate as stage pipelines: an LZ-style parser identifies repetitions; a probability model estimates symbol/event likelihoods; and an entropy coder (e.g., Huffman, ANS, range coder) emits the bitstream. Practical implementations extend this basic pipeline with guard rails and fallback stages---e.g., raw/uncompressed blocks for incompressible regions (DEFLATE) and RLE blocks for degenerate runs or one-byte blocks (Zstandard)---and can bypass encoding for tiny inputs where headers would dominate.

This staged idea is explicit in several widely used formats. \textbf{PNG}~\cite{rfc2083} applies a per-scanline prediction filter (None/Sub/Up/Average/Paeth) before DEFLATE, turning image structure into locally predictable residuals that the downstream coder compresses well; the filter choice is itself a stage decision recorded per row. \textbf{Blosc}~\cite{BLOSC} composes blocking, a shuffle/bitshuffle transform (to decorrelate bytes/bits and surface runs), and then a configurable backend codec (LZ4, Zstd, etc.), executed in a multithreaded pipeline to maximize memory locality and throughput.

Some systems make the composition selectable. Blosc assembles a per-chunk, linear pipeline---e.g., \emph{(shuffle\,|\,bitshuffle\,|\,delta\,|\,trunc\_prec) $\to$ codec (LZ4, Zstd, \ldots{})}---with knobs for block size and parallelism. \textbf{Parquet}~\cite{PARQUET} composes per-column encodings (dictionary, delta, RLE/bit-packing) with a backend codec. \textbf{ZPAQ} goes further: the archive stores a virtual-machine program~\cite{mahoney2015zpaq} specifying contexts/transforms and the coder. Here, the pipeline is part of the compressed frame. \textbf{BTune}~\cite{BTUNE} explores automatic choice/parameter search across Blosc2 codecs and filters.

In practice, these designs are constrained by enumerated stage catalogs and fixed rules on how these stages can be composed. Even where plugins exist, tuning often focuses on parameters for individual stages rather than exploring different stage orderings or richer graphs. As a result, a large fraction of the transform design space---and the cross-stage interactions that dictate effectiveness---remains practically under-explored.

\section{Core Concepts}
\label{sections:concepts}

\NAME{} represents the natural evolution of the staged pipeline design. Rather than limit the system to a strict linear flow, we introduce the graph model of compression. In addition to branching and dynamism, this new framing unlocks the possibility of training by imposing structure onto the process of compression.

\subsection{Data and Messages}

In typical data compression parlance, a \textit{message} is a sequence of bytes. Formally, we denote this
$$\mu \in \mathcal{B}^*$$
where $\mu$ is the message and $\mathcal{B} = \{\verb|0x00|, \verb|0x01|, \dots , \verb|0xff|\}$ is the set of 8-bit bytes.

In the graph model, we adopt a stricter requirement for messages.

\begin{definition}[Message Sets]
    A \textbf{message set} is a non-empty subset of the universe of bitstrings.
\end{definition}

Under this framing, a \textit{message} is an element drawn from a message set. Rather than be any random bitstring, we impose semantic requirements on messages by restricting the possibility set. For instance, components may require that messages represent 64-bit integer arrays by requiring that all messages have bit-length divisible by 64. More restrictive requirements can also be imposed, like requiring all messages in the set to be sorted runs of bytes.

\subsection{Codecs}

Fundamentally, a codec is just a function operating on message sets.

\begin{definition}[Codec]
An input (resp. output) is an ordered tuple of messages $\mu = (\mu_1, \dots, \mu_n)$, each drawn from potentially different message sets $\mu_i \in X_i$. The \textbf{input domain} (resp. \textbf{output domain}) is the ordered tuple of these message sets, i.e. $X = (X_1, \dots, X_n)$.
A \textbf{codec} is a tuple $(C,D)$ of functions. The \textit{encoder} $C: I \to O$ is a mapping between a non-empty input domain $I$ and a non-empty output domain $O$. The \textit{decoder} $D: O \to I'$ maps $O$ to a possibly different regenerated domain $I'$. A codec is \textit{lossless} if this mapping is invertible, that is, $I \equiv I'$ and
$$D(C(\mu)) \equiv \mu, \; \forall \mu \in I \;.$$
\end{definition}

This definition intentionally de-emphasizes the inner workings of the codec. In our model, the input/output signature matters more than the exact implementation because the signature tells us how the information is transformed semantically. This is an important abstraction, as it enables us to consider composition at a higher level.

\begin{remark}
    From an implementation perspective, codecs are most useful when they are small and limited in scope. \Cref{sections:implementation} describes the implementation philosophy in \NAME{}.
\end{remark}

\subsection{Composition and Graphs}

In the graph model, compressors are graphs built from codec nodes. As a motivating example, consider the \texttt{tokenize} codec. Briefly, \texttt{tokenize} searches for repeated instances of the same ``token''. It works by taking a message $\mu \in \Sigma^*$ and outputting 2 messages: $\alpha$, the list of unique tokens in $\mu$; and $\nu$, the ``index'' in $\alpha$ of each token within $\mu$.

\begin{figure}[H]
    \centering
    \footnotesize
    \begin{tikzpicture}[]
    \node [shape=rectangle,fill=orange!20, align=center](table1) at (0.0,0.0) {
            $\mu$ \\
            \begin{tabular}{c|c|c|c|c|c|c} \toprule
                \verb|alice| & \verb|bob| & \verb|bob| & \verb|eve| & \verb|alice| & \verb|bob| & \verb|alice|\\
                \bottomrule
            \end{tabular}
        };
    \node [shape=rectangle,fill=blue!20, align=center](table2) at (-1.0,-2.0) {
            $\nu$ \\
            \begin{tabular}{c|c|c|c|c|c|c} \toprule
                0 & 1 & 1 & 2 & 0 & 1 & 0\\
                \bottomrule
            \end{tabular}
        };
    \node [shape=rectangle,fill=blue!20, align=center](table3) at (3.0,-2.0) {
            $\alpha$ \\
            \begin{tabular}{c} \toprule
                \verb|alice|\\ \midrule
                \verb|bob|\\ \midrule
                \verb|eve|\\
                \bottomrule
            \end{tabular}
        };
    \draw [thin, -Stealth] (table1) -- (table2);
    \draw [thin, -Stealth] (table1) -- (table3);
\end{tikzpicture}
    \caption{An example invocation of the \texttt{tokenize} codec.}
    \label{fig:tokenize}
\end{figure}
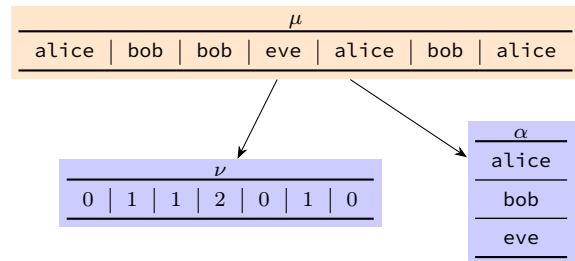

Tokenization is sometimes an effective compressor on its own (such as when the message is composed of many repetitions of a few large tokens). More frequently though, its utility is as an intermediate transformation, which produces outputs better suited for subsequent processing. We can attach other codecs to each of the \texttt{tokenize} codec's two outputs, which can separately attack the problems of efficiently representing the contents of the tokens and the indices.

\begin{figure}[H]
    \centering
    \footnotesize
    \begin{tikzpicture}[]
    \node [shape=rectangle, draw, fill=white, align=center](ROOT) at (2.0,1.5) {};
    \node [shape=ellipse, draw, fill=white, align=center](TOK) at (2.0,0.0) {
        \verb|tokenize|
    };
    \node [shape=ellipse, draw, fill=white, align=center](HUFF) at (0.0,-1.5) {
            \verb|huffman|
    };
    \node [shape=ellipse, draw, fill=white, align=center](LZ77) at (4.0,-1.5) {
        \verb|LZ77|
    };
    \node [shape=rectangle, draw, fill=white, align=center](store1) at (0.0, -2.5) {};
    \node [shape=rectangle, draw, fill=white, align=center](store2) at (4.0, -2.5) {};
    \draw [thin, -Stealth] (ROOT) -- (TOK) node [midway, fill=orange!20] {$\mu$};
    \draw [thin, -Stealth] (TOK) -- (HUFF) node [midway, fill=blue!20] {$\nu$};
    \draw [thin, -Stealth] (TOK) -- (LZ77) node [midway, fill=blue!20] {$\alpha$};
    \draw [thin, -Stealth] (HUFF) -- (store1);
    \draw [thin, -Stealth] (LZ77) -- (store2);
\end{tikzpicture}
    \caption{An example compressor that uses tokenize, Huffman, and LZ77.}
    \label{fig:tokenize_compressor}
\end{figure}
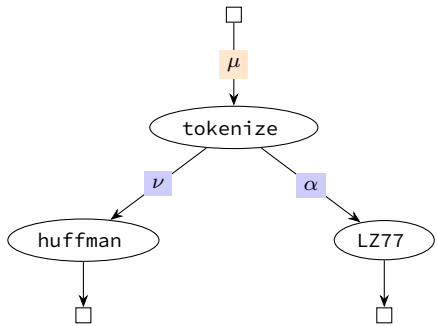

While the alphabet $\alpha$ has the same type of content as the original message $\mu$, the indices in $\nu$ are a sequence of integers rather than strings. While $\nu$ is a partial representation of $\mu$, by transforming it into a fixed-width, integer sequence, we can bring techniques to bear on it that can't be applied directly to $\mu$. For instance, we may construct a compressor that sends $\alpha$ to an LZ77 compressor and $\nu$ to an entropy encoder like Huffman. \Cref{fig:tokenize_compressor} contains a visualization of this new compressor.

As the visualization implies, a compressor with multiple codecs conveniently organizes itself as a graph. Informally, a \textbf{compression graph} is a computational graph~\cite{COLLINS18, DYER16} where the nodes represent codecs and edges represent input and output sets\footnote{Technically, this is a \emph{reversed} computational graph, since in typical depictions the feed-forward direction merges multiple inputs to produce the output, whereas a compression graph generates multiple outputs from the input.} In particular, an edge between a parent and child codec indicates a ``happens-after'' relationship, where one of the outputs of the parent is used as one of the inputs of the child.

\begin{definition}[Compression Graph]
    A \textbf{computational graph} is a directed, acyclic, graph (DAG) where the nodes are functions and the edges represent function arguments (and data dependencies).
    A \textbf{compression graph} is a computational graph where each node $v$ is labelled with a codec $C_v: I_v \to O_v$, and edges $u \to v$ are doubly-labelled with both an output from the source $(O_u)_i$ and an input to the target $(I_v)_j$ such that $(O_u)_i \subseteq (I_v)_j$.
\end{definition}

The sequence of transforms permitted by this model allows us to build compressors that exploit the semantics of the data much better than generic compressors. Semantic specialization in intermediate streams increases as you traverse the compression graph, increasing the efficiency of subsequent codecs. For entropy coders, such specialization can result in intermediate representations with lower entropy and thus a more compact code.

\subsection{Universal Decompression}

The compression graph inherits some useful properties from computational graphs. Notably, computational graphs are DAGs. And since every DAG admits a topological sort, this ensures that a well-defined compression graph always admits a valid feed-forward computation order (for compression) and a valid backpropagation order (for decompression). Moreover, the proper decode procedure is uniquely determined just by knowing the ultimate outputs and the graph structure.

So long as each codec has a well-defined inverse, the DAG structure allows us to treat decompression as a procedural exercise. \NAME{} exploits this property to provide a universal decompressor.

\subsection{Runtime Dynamicity}

Strong guarantees on decodability allow for more flexibility on the compression side. Indeed, so long as a compressor generates a compression graph \emph{somehow}, the compressed frame will be decodable. Finding an exhaustive list of rules for dynamicity within the graph model is likely a hard problem. In this section, we propose one such source of dynamicity by extending the graph model slightly.

\begin{definition}[Function graph]
    Denote by $\mathbb{G}$ the set of compression graphs. A \textbf{function graph} is a function $F: I \to \mathbb{G}$. A \textbf{dynamic compression graph} is a compression graph where the nodes are either codecs or function graphs.
\end{definition}

\begin{figure}[h]
    \centering
    \footnotesize
    \begin{tikzpicture}[
  node/.style={circle, draw},
  edge/.style={-Stealth},
]

\node[node,                      ] (A1) {};
\node[node, below left =0.5 of A1] (B1) {};
\node[node, below right=0.5 of A1] (C1) {};
\node[node, below left =0.5 of B1] (D1) {};
\node[node, below right=0.5 of B1] (E1) {};
\node[node, below right=0.5 of C1, fill=Blue] (F1) {};

\draw[edge] (A1) -- (B1);
\draw[edge] (A1) -- (C1);
\draw[edge] (B1) -- (D1);
\draw[edge] (B1) -- (E1);
\draw[edge] (C1) -- (F1);

\node[node, below      =1.75 of A1] (A2) {};
\node[node, below left =0.5  of A2] (B2) {};
\node[node, below right=0.5  of A2] (C2) {};
\node[node, below left =0.5  of B2] (D2) {};
\node[node, below right=0.5  of B2] (E2) {};
\node[node, below right=0.5  of C2, fill=Dandelion!25] (F2) {};
\node[node, below left =0.5  of F2, fill=Blue] (G2) {};
\node[node, below right=0.5  of F2, fill=Dandelion!25] (H2) {};
\node[node, below      =0.5  of G2, fill=Dandelion!25] (I2) {};
\node[node, below      =0.5  of H2, fill=Dandelion!25] (J2) {};
\draw[edge] (A2) -- (B2);
\draw[edge] (A2) -- (C2);
\draw[edge] (B2) -- (D2);
\draw[edge] (B2) -- (E2);
\draw[edge] (C2) -- (F2);
\draw[edge] (F2) -- (G2);
\draw[edge] (F2) -- (H2);
\draw[edge] (G2) -- (I2);
\draw[edge] (H2) -- (J2);

\node [draw, thick, dotted, fit=(F2) (G2) (H2) (I2) (J2), inner sep=2pt] (EXPANSION2) {};

\draw [decorate,decoration={brace,amplitude=0.25cm}]
      (EXPANSION2.north east)
      -- node [midway] (EXPANSION2TOP) {}
      (EXPANSION2.south east);

\draw [densely dashed] (F1) to (EXPANSION2TOP|-F1) to [out=  0, in=  0] ($(EXPANSION2TOP)+(0.25,0)$);

\node[node, below      =3   of A2] (A3) {};
\node[node, below left =0.5 of A3] (B3) {};
\node[node, below right=0.5 of A3] (C3) {};
\node[node, below left =0.5 of B3] (D3) {};
\node[node, below right=0.5 of B3] (E3) {};
\node[node, below right=0.5 of C3] (F3) {};
\node[node, below left =0.5 of F3, fill=Dandelion!25] (G3) {};
\node[node, below right=0.5 of F3] (H3) {};
\node[node, below left =0.5 of G3, fill=Dandelion!25] (I3) {};
\node[node, below right=0.5 of G3, fill=Dandelion!25] (J3) {};
\node[node, below right=0.5 of I3, fill=Dandelion!25] (K3) {};
\node[node, below      =0.5 of K3] (L3) {};
\node[node, below      =0.5 of H3] (M3) {};
\draw[edge] (A3) -- (B3);
\draw[edge] (A3) -- (C3);
\draw[edge] (B3) -- (D3);
\draw[edge] (B3) -- (E3);
\draw[edge] (C3) -- (F3);
\draw[edge] (F3) -- (G3);
\draw[edge] (F3) -- (H3);
\draw[edge] (G3) -- (I3);
\draw[edge] (G3) -- (J3);
\draw[edge] (I3) -- (K3);
\draw[edge] (J3) -- (K3);
\draw[edge] (K3) -- (L3);
\draw[edge] (H3) -- (M3);

\node [draw, thick, dotted, fit=(G3) (I3) (J3) (K3), inner sep=2pt] (EXPANSION3) {};

\draw [decorate,decoration={brace,amplitude=0.25cm}]
      (EXPANSION3.south west)
      -- node [midway] (EXPANSION3TOP) {}
      (EXPANSION3.north west);

\draw [densely dashed] (G2) to (EXPANSION3TOP|-G2) to [out=180, in=180] ($(EXPANSION3TOP)+(-0.25,0)$);

\end{tikzpicture}
    \caption{An example of function graph expansion. Function graphs are shaded and their expansions marked in dotted lines. }
    \label{fig:function_graph}
\end{figure}
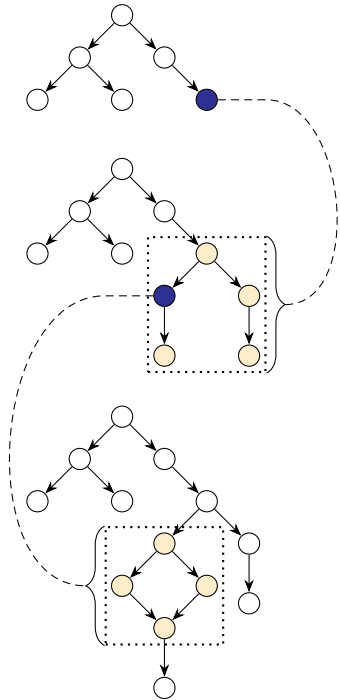

Function graphs can be described as ``selectors'', choosing a graph based on the input, which itself may contain additional function graphs. At runtime, this expansion naturally modifies the graph being run. \Cref{fig:function_graph} illustrates this process.

\subsection{Graph Resolution}

Readers familiar with lambda calculus may draw a vague parallel between function graph expansion and beta-reduction. Like beta-reduction, the function graph expansion process creates another valid graph, which may have more opportunities for function graph expansion. The ``beta-normal form'' for graphs is similarly important.

\begin{definition}[Resolved Graph]
    A \textbf{resolved graph} is a compression graph that contains no function graphs.
\end{definition}

The example graph in \cref{fig:tokenize_compressor} is by definition a resolved graph, as are all graphs with no dynamism. Furthermore, a compression that succeeds will always generate a resolved graph. Since the resolved graph contains only regular codecs, it also completely specifies how to reconstruct the original input. In the graph model, the decoder cannot make any runtime decisions based on data presented, so the existence of a resolved graph allows us to incorporate dynamism into \NAME{}.

\subsection{Practical Implications}

The graph model enables two new freedoms that address pain points common to application-specific compression:

\begin{itemize}
  \item \textbf{Specialization without Compromising Support.} Typically, specialization is always a tradeoff; one must weigh the benefits against the obvious drawbacks of less troubleshooting support and a sparser tooling environment. However, the universality of graph compression means domain experts can develop improved compression strategies for specific data while maintaining access to the \NAME{} ecosystem and the tools built for it.
  \item \textbf{Constraint-Aware Tradeoffs.} The flexibility to build multiple configs for the same data unlocks additional configuration opportunities. One of these is the tradeoff between speed and compression ratio. By simply changing the graph chosen at runtime, you can shift your usage along this tradeoff curve.
\end{itemize}

The delegation of decision-making to compression-time is an important property of \NAME{}'s implementation of the graph model. By separating the decision-making process from the on-disk representation, progress in compression theory can be made while maintaining strict stability of the wire format. In particular, offline training of compressors is enabled by this split. This is a generalization of the idea of dictionary training and is developed further in \cref{subsections:building_a_compressor_tools}, \cref{subsec:training}, and \cref{subsections:training}.

\section{Building Compressors}
\label{sections:building_a_compressor}

The freedom afforded by \NAME{}'s graph model is a double-edged sword: the capability it offers comes at the cost of a combinatorial explosion of choices that must be made---namely which components to use and how to compose and configure them. There is no one-size-fits-all compressor. However, in practice, many \NAME{} compressors do share the same overall structure:

\begin{figure}[H]
    \centering
    \begin{tikzpicture}
\node [draw, fill=white] (PARSE) {Parse\vphantom{jk}};
\node [draw, fill=white, right=0.375 of PARSE] (GROUP) {Group\vphantom{jk}};
\node [draw, fill=white, right=0.375 of GROUP] (TRANSFORM) {Transform\vphantom{jk}};
\node [draw, fill=white, right=0.375 of TRANSFORM] (COMPRESS) {Compress\vphantom{jk}};

\draw [->] ($(PARSE.west)+(-0.375,0)$) -- (PARSE);
\draw [->] (PARSE) -- (GROUP);
\draw [->] (GROUP) -- (TRANSFORM);
\draw [->] (TRANSFORM) -- (COMPRESS);
\draw [->] (COMPRESS) -- ($(COMPRESS.east)+(0.375,0)$);

\draw [decorate, decoration={brace,amplitude=5pt,raise=0.0625cm}]
      (PARSE.north west)
      -- node [midway, above=5pt] {\scriptsize Frontend}
      (GROUP.north east);

\draw [decorate, decoration={brace,amplitude=5pt,raise=0.0625cm}]
      (COMPRESS.south east)
      -- node [midway, below=5pt] {\scriptsize Backend}
      (TRANSFORM.south west);
\end{tikzpicture}
    \caption{Common abstract compressor structure.}
\end{figure}
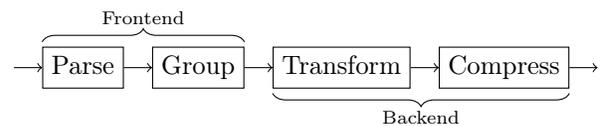

This pattern emerges because the components of \NAME{} that are actually good at compressing data---its suite of transforms and compressors---work best on homogenous streams of data. For inputs that aren't already organized that way, those backend components require a frontend to parse and group the input into streams, which the backend can then compress effectively.

\subsection{A Motivating Example}

Consider a CSV file. The layout of a CSV file is a row-major representation of the data, which packs data of different types next to each other. A CSV parser for \NAME{} would likely want to organize the input data by column instead of row, by creating an output stream for each column and sending all the values in that column to that output (and additionally creating an output for the framing commas and newlines). Reshaping and separating the input this way puts the data in a column-major representation, with streams of like-typed data uninterrupted by distractions (commas, whitespace, other values, etc.). This structure is likely to significantly improve \NAME{}'s ability to operate on the data effectively. (And in fact, this is exactly how the CSV parser bundled in \NAME{} works.)

However, this splitting is likely to be too fine-grained, because it has mechanically partitioned every single component of the input, even when they might have significant exploitable correlations. This parse has traded away the ability to use cross-column correlations in order to be able to use cross-row correlations. To address this, we can send the parser's outputs to a grouping stage, which searches for cross-stream correlations and re-groups them where useful.

After parsing and grouping, the outputs are ready to be sent to \NAME{}'s backend graphs. These graphs generally begin with some number of transformations and end in compression. Loosely, the former try to change the representation of data with the goal of making it more compressible (e.g., parsing ASCII integers to a binary representation, or applying a linear predictor to a stream of values), while the latter take advantage of the structure and regularity that is exposed by the preceding stages to produce a compact representation (e.g., LZ compression and entropy coding).

\subsection{Parsing Stage}
\label{subsections:parsing_stage}

The job of the parsing stage is to take an input stream and separate the data into its logical components. Since \NAME{} is a lossless compressor, every byte in the input must be reconstructible from the parsed outputs. This parse is typically achieved by providing a parsing function to \NAME{} which can be plugged into one of the standard \textbf{dispatch} codecs. These codecs apply the instructions produced by the parsing function describing how to dispatch each byte of the input stream into a set of output streams. Alternatively, the \textbf{Simple Data Description Language (SDDL)}, described in \cref{subsec:sddl}, can implement the parsing function for you, given an SDDL description of the format. Because the dispatch codecs record the instructions they're given into their output streams, the parsing logic specific to the file format in question is needed only on the compression side and \NAME{}'s universal decoder can decompress and reconstruct the original data without any external instruction about the file format. 

Some example parsing strategies we have implemented:

\begin{itemize}
    \item In vector data, like PyTorch tensors or {\tt WAV} audio files, the interesting contents of the vector (the actual weights/features/samples) are extracted into their own output, separate from the framing, control, and meta-data.
    \item In tabular data structures like CSV and Parquet, each column is parsed into its own output, and there is an additional output for the framing, control, and meta-data.
    \item In tree-shaped data structures, like Thrift and Protobuf, each unique path is parsed into its own output, and there are a couple additional outputs for the framing and control data.
\end{itemize}

This stage usually represents the bulk of the work of writing an \NAME{} compressor for a new format, since this is the piece that \NAME{} cannot automate. Once the structure and semantics of the input have been captured via this parsing process, the subsequent stages are much more amenable to automation, and \NAME{} provides tooling to do so.

Note that a parsing stage isn't always necessary. This is trivially the case when the data in question is already in the format \NAME{}'s backends desire (a clean, linear vector of data). In addition though, while parsing allows \NAME{} to operate effectively on existing serialization formats, an alternative integration model skips this. Unlike traditional compressors which are limited to operating on a single bytestream, \NAME{} accepts multiple, typed input streams to compress. This enables a more direct integration that skips over both serializing the structured data into a linear bytestream and the immediate reversal of that in parsing. Instead users can marshal their data directly into a set of typed buffers and pass that to \NAME{}. Not only does this lessen the work \NAME{} has to do, it can save a meaningful amount of serialization and deserialization work outside of \NAME{}.

\subsection{Grouping Stage}
\label{subsections:grouping_stage}

The parsing stage often breaks the input up into units that are too fine-grained. In order to take advantage of correlation between these streams they must be grouped back together. For example, that grouping can be done by concatenating or interleaving streams.

The parser could simply not split the data up this finely, but we've found that separating the concern of grouping out of the parser both simplifies the parsing logic, and allows us to share the grouping logic between compressors for all formats.

\NAME{} provides a clustering graph and corresponding trainer that handles this stage. Given samples of the parsed data, it searches for correlation and determines how streams should be grouped together.

\subsection{Transformation Stage}
\label{subsections:transformation_stage}

The transformation stage applies any domain-specific transformations to the data. This is where compressor authors can leverage their understanding of the data to drastically improve compression. For example, the \texttt{delta} codec can be applied to sorted integers. Or the \texttt{parse\_int} codec can be applied to convert ASCII integer strings to integers. Intelligent transformation choice often leads to compression that is both faster and stronger.

While compressor authors can determine their own transformations, \NAME{} provides tooling to make this stage easier. First, the generic compression backends detect common patterns and handle them effectively. Second, \NAME{} provides the \textbf{Automatic Compression Explorer (ACE)} (described in \cref{subsec:training}), which determines the best backend graph for a given input.

\subsection{Compression Stage}

Finally, the compression stage runs generic compression after all domain-specific knowledge has been leveraged. Compressor authors are expected to simply use \NAME{}'s builtin backend compressors, or use the backend graph produced by ACE.

The backend component is typically a LZ codec or an entropy codec. \NAME{} provides generic backend graphs for these purposes: \textbf{Compress} and \textbf{Entropy}. These backends select the right compression or entropy engine for the data.

In addition to the \texttt{zstd} LZ backend, \NAME{} provides the \texttt{field\_lz} codec and builtin backend graph. This codec operates on struct streams rather than bytes, which allows for more efficient LZ compression of numeric and struct streams.

\subsection{Construction Strategies}
\label{subsections:building_a_compressor_tools}

Even if a compressor conforms to this overall structure, figuring out how to best group, transform, and compress a dataset can be a non-trivial exercise. There are several approaches to making these decisions:

\begin{description}
\item[Manually] The user can explicitly dictate the structure and configuration of their compressor, based on intuition or experimentation. This is how application-specific compressors have been built historically. This approach has several downsides. It is laborious and difficult and we have frequently found (to our surprise) that it produces sub-optimal results.

\item[Runtime Automatic Experimentation.] The runtime dynamicity offered by OpenZL allows experimentation with multiple processing options to occur at any point in the flow of compression. Results can be observed, the best option can be selected and the corresponding behavior can be executed. While possible for highly asymmetric scenarios, na\"ive brute-force approaches to exploration are impractically slow for most compression workflows, and even very constrained versions of this approach are likely to impose significant compression speed costs.

\item[Offline Automatic Experimentation.] When tasked with compressing a large, relatively homogeneous dataset, a user might want to pick a small, representative sample corpus. The same automated exploration of options described previously, which would be too slow to run practically during each compression, may be more feasible when run on the much smaller selected subset of the data. The results of that exploration can be collected and used in the compression of the dataset as a whole.
\end{description}

While \NAME{} supports all of these strategies, the last in particular has proven particularly effective. \NAME{} has accumulated several tools that aim to optimize a compressor to compress a given corpus, unified into a \textbf{training} workflow described in \cref{subsec:training}.

\section{Implementation}
\label{sections:implementation}

This section provides an overview and rationale for the structure of the current implementation of \NAME{}. 

\subsection{The Software Stack}

The open-source OpenZL implementation is layered, with the goal of making the hot path efficient and verifiable.
At the bottom sits the C11 core library, \texttt{libopenzl}, which exposes a stable surface and an execution engine for compression graphs.
A thin C++ façade wraps that surface to provide RAII and strong typing. On top, a Python binding offers an API that integrates with data-science toolchains while preserving the core library's determinism and memory discipline.
Two tools complete the stack: \textit{SDDL} (\cref{subsec:sddl}), which turns a format description into a parser configuration aligned with the core type system, and the \textit{trainer} (\cref{subsec:training}), which produces serialized compressor configurations from samples.

The choice of C11 for the core is deliberate: The wide portability and ubiquity of C toolchains, predictable memory semantics, universal ABI, and clear debuggability expectations  are critical at this layer. By constraining the core to a narrow contract, higher layers can evolve at their own cadence without perturbing the on-wire format or the decoder.

\subsubsection{Public APIs: C++ and Python}

The C++ layer is intentionally thin. It wraps the C handles to manage resource lifetimes, translate error codes to exceptions, and make it easier to write custom components. Its primary role is to make core primitives easy to use in larger C++ systems, while still allowing a drop down to C where necessary.

The Python bindings mirror the C++ bindings closely. They focus on efficiency, allow zero-copy interaction with OpenZL through the buffer protocol, and can expose buffers as NumPy arrays, PyTorch tensors, \texttt{dlpack} tensors, or \texttt{bytes}. In addition to enabling production usage of OpenZL in Python, Python notebooks are a convenient way to experiment with custom OpenZL compressors on new datasets. The ability to write custom components in Python speeds up the prototyping process.

\subsection{Implementation Details}

\NAME{}'s implementation of the graph model is mostly straightforward. We mention some salient details here.
\begin{description}
    \item[Codecs.] Codecs are the lowest level in the OpenZL architecture. Each codec does one thing well. Codecs are split into two parts: \emph{encoder} and \emph{decoder}. Implementation-wise, each side is typically organized into two layers: a \emph{kernel} and a \emph{binding}. Kernels are small, deterministic, and allocation free; the binding around them handles types, bounds, and buffers. The vast majority of CPU time is spent in the kernel, so splitting in this way simplifies performance optimization work.
    \item[Message Sets.] \NAME{} has a partial implementation of message sets. It would be unrealistic to allow specifying arbitrarily-specific sets, so we approximate it with a type system. There are currently 4 types:
    \begin{itemize}
        \item \texttt{bytes} for opaque serial data.
        \item \texttt{string} for sequences of byte strings.
        \item \texttt{struct(k)} for fixed-size (\(k \ge 1\)) records.
        \item \texttt{numeric(w)} a specialization of \texttt{struct} for host-endian 8, 16, 32, and 64-bit numbers.
    \end{itemize}
    \item[Dynamism.] We implement function graphs in two ways. The \textit{selector} is a faithful representation of the abstract definition. All it does is output a graph based on input data. We found an additional construct useful. Named in the code as ``function graphs'', these are regular codec encoders with the restriction that they cannot modify the input data, only call other codecs to do what they want. This is a happy medium between implementation power and abstract correctness.
\end{description}

\subsection{Versioning and Decoding}

An OpenZL compression session is driven by the compressor config. It targets a particular decoder profile, which describes---among other things---a specific format version. This makes it possible to enforce compatibility in an ecosystem with multiple generations of decoders.
Based on the selected policy, it produces a \textit{compressed frame}, or \textit{frame} for short. The frame header declares the format version and capability requirements, that the decoder can check against its profile.

A frame is organized into \textit{chunks}, which can be decoded independently.
Each chunk starts with a description of its resolved graph followed by the edge payloads (leaf data) and, optionally, integrity checksums. This is enough for the decoder to safely rebuild the original payload without sideband information.

\subsection{SDDL: A Parser Builder}
\label{subsec:sddl}

\Cref{subsections:parsing_stage} describes several strategies to tackle the problem of getting data into a form on which \NAME{} can operate effectively. While fancier integration options have their benefits, the non-invasive baseline is to teach \NAME{} how to parse the data in its existing format.

Rather than writing and plugging in a parsing function as a custom component to \NAME{}, \NAME{} offers the \textbf{Simple Data Description Language (SDDL)}, which allows users to write a textual description of the format of their data. The SDDL engine applies this description to the input, using it to identify the tokens that make up the input and then to dispatch them into typed output streams.

SDDL is composed of several components: a compiler which translates the text description into a bytecode, an execution engine that applies that bytecode to an input, producing a stream of dispatch instructions, and the dispatch codecs which actually perform the decomposition of the input.

SDDL's execution engine offers a sandboxed runtime which mitigates the security and stability pitfalls typical of custom parsing logic. Despite these constraints, SDDL still offers dynamicity, which means descriptions can describe \emph{formats} rather than \emph{specific inputs}.

\subsection{Training}
\label{subsec:training}

The \NAME{} trainer implements the offline experimentation strategy presented in \cref{subsections:building_a_compressor_tools} for constructing a compressor. Given a corpus of sample inputs and a seed compressor, the trainer mutates the compressor in search of configurations that give better performance on the given dataset.

The trainer exploits several convenient properties of \NAME{}:
\begin{description}
    \item[Abstract Structure.] \NAME{}'s support for the inspection and manipulation of graphs in the abstract facilitates the development of generic algorithms to generate and evaluate candidate compressors.
    
    \item[Serializable Compressors.] Because \NAME{} compressors are serializable, training can be a standalone process. The compressor configuration that results from training can be published to the use case and adopted seamlessly, without rebuilding or restarting applications.
\end{description}

\begin{figure}[H]
    \centering
    \begin{tikzpicture}
    \node [draw, fill=white] (INPUT1) {Inputs};
    \node [draw, fill=white, below right=0.0625 of INPUT1.center, anchor=center] (INPUT2) {Inputs};
    \node [draw, fill=white, below right=0.0625 of INPUT2.center, anchor=center] (INPUT3) {Inputs};
    \node [draw, fill=white, below right=0.0625 of INPUT3.center, anchor=center] (INPUTS) {Inputs};
    
    \node [draw, fill=white, right=3.5 of INPUT1] (OUTPUT1) {Outputs};
    \node [draw, fill=white, below right=0.0625 of OUTPUT1.center, anchor=center] (OUTPUT2) {Outputs};
    \node [draw, fill=white, below right=0.0625 of OUTPUT2.center, anchor=center] (OUTPUT3) {Outputs};
    \node [draw, fill=white, below right=0.0625 of OUTPUT3.center, anchor=center] (OUTPUTS) {Outputs};
    
    \node [below left=0.1875 and 1 of INPUTS, inner sep=0pt] (LINE_LEFT) {};
    \node [below right=0.1875 and 0.5 of OUTPUTS, inner sep=0pt] (LINE_RIGHT) {};
    
    \draw [dotted] (LINE_LEFT) -- (LINE_RIGHT);
    
    \node [txt, tiny, above right=0.0625 and 0 of LINE_LEFT, inner sep=0pt] {online};
    \node [txt, tiny, below right=0.0625 and 0 of LINE_LEFT, inner sep=0pt] {offline};
    
    \coordinate (MID) at ($(INPUT1)!.5!(OUTPUT1)$);
    
    \node [draw, fill=white, right=0.125 of MID, anchor=center] (COMPRESSOR1) {Compressions};
    \node [draw, fill=white, below right=0.0625 of COMPRESSOR1.center, anchor=center] (COMPRESSOR2) {Compressions};
    \node [draw, fill=white, below right=0.0625 of COMPRESSOR2.center, anchor=center] (COMPRESSOR3) {Compressions};
    \node [draw, fill=white, below right=0.0625 of COMPRESSOR3.center, anchor=center] (COMPRESSORS) {Compressions};
    
    \draw [->] (INPUTS) -- (COMPRESSOR1.west|-COMPRESSORS.west);
    
    \draw [->] (COMPRESSORS) -- (OUTPUT1.west|-OUTPUTS.west);
    
    \coordinate (TRAINER_MID) at ($(INPUTS)!.5!(COMPRESSORS)$);
    
    \node [draw, fill=white, below=1 of TRAINER_MID, anchor=center] (TRAINER) {Trainer};
    
    \draw [->]
          (INPUTS)
          -- node [txt, tiny, pos=0.67, fill=white, inner sep=0pt] {samples}
          ([yshift=0.075cm]INPUTS|-TRAINER)
          --
          ([yshift=0.075cm]TRAINER.west);
    \draw [->]
          (TRAINER)
          --
          (COMPRESSORS|-TRAINER)
          -- node [txt, tiny, pos=0.33, fill=white, inner sep=0pt] {config}
          (COMPRESSORS);
    
    \node [txt, tiny, inner sep=0pt] at ($(TRAINER.west)+(-1.25,-.075)$) (PROFILE) {compressor};
    \draw [->] (PROFILE) -- ($(TRAINER.west)+(0,-.075)$);
\end{tikzpicture}
    \caption{Abstract compressor training workflow.}
\end{figure}
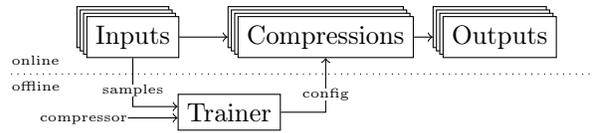

\subsubsection{Orchestration}

The trainer consists of an orchestration framework and a registry of training plug-ins. The registry associates configurable codecs and function graphs with corresponding logic to train them. During training, the orchestrator executes the existing compressor on the sample corpus until it reaches a component which has a registered training function, which it invokes. The component's training function performs local experimentation and uses that to parameterize or restructure the component in question and its successors. After applying these changes, the orchestration framework resumes the execution of the sample compressions, with the new compressor structure, and continues traversing the compression graph until it encounters the next trainable component.

Individual component trainers can also return multiple configuration options (at different points in the speed/ratio tradeoff space). These options allow the orchestration framework to make globally-efficient choices concerning where to allocate compression and decompression time across the whole compressor.

In this way, specialized, local optimization logic can be composed into a holistic compressor optimizer\footnote{Note that the forward pass design described here does suffer from a limitation: components can have cyclic training dependencies. This remains an area of active experimentation.}.

\subsubsection{Components}

This section describes some specific training components that have demonstrated value:

\begin{description}
    \item[Clustering Trainer.] The \textbf{clustering trainer} is responsible for optimizing the ``grouping'' stage of the compressor. It partitions the input into fine-grained units, then picks a grouping for these units in a way that optimizes a given cost function---typically compressed size. The space of possible groupings of inputs is large, and evaluating compressed size is essentially a black box function, so we combine heuristic search with training to find a good grouping.
    
    \item[Backend Generation.] The \textbf{Automated Compressor Explorer (ACE)} uses a genetic algorithm to compose codecs together into backend graphs. As part of its exploration of possible graphs, it attempts to produce a maximally-wide range of options across the performance trade-off frontier.
    
    \item[Selector Training.] Classification is a classic machine learning problem. The \textbf{ML Selector} applies this to graph selection by using statistical features extracted from the data to choose the best graph for a given input. The trainer builds the model that the selector uses for inference during compression.
\end{description}

\section{Experimental Results}
\label{sections:results}

In this section, we demonstrate the flexibility and efficacy of \NAME{} by building compressors for a wide variety of datasets and data formats. \Cref{subsections:ratio-results} proves the viability of the graph model by finding high-compression ratio compressors for a variety of benchmark datasets. \Cref{subsections:pareto-frontier} visualizes the Pareto frontier of \NAME{}'s trained compressors on the axes of compression ratio and compression speed. We finish with \cref{subsections:enwik}, a case study showing some of the current limitations of the system.

\subsection{Hardware and Software}
All benchmarks were run on a Lenovo P620 desktop with an AMD Ryzen Threadripper PRO 3995WX CPU, with 256 GB of memory (8$\times$32 GB DDR4 3200MHz RDIMM ECC memory), and a 2 TB Samsung PM981a SSD\footnote{Swap was not used. All benchmarks operate in memory.}. Precision Boost was disabled for consistent results.

We compiled \NAME{} with GCC~14 on Fedora Linux~41. Each dataset is also benchmarked against a list of widely-used general compressors. We use \texttt{lzbench}, an open-source benchmarking tool for open-source compressors~\cite{LZBENCH}, for the measurement. We chose \texttt{xz}, \texttt{zstd}, and \texttt{gzip}. \Cref{table:generic_compressors} summarizes some of their key properties. For some comparisons, we also compared against custom compressors. These are detailed in their respective sections.

\begin{table*}[ht]
    \centering
    \begin{NiceTabular}{ l l S S }
    \CodeBefore
    \Body
    \toprule
    \multicolumn{1}{c}{Compressor} & \multicolumn{1}{c}{Strategy} & \multicolumn{1}{c}{Min Level} & \multicolumn{1}{c}{Max Level} \\
    \midrule
    \verb|xz| & LZ77 + Context Model + Range & 1 & 9\\
    \verb|zstd| & LZ77 + Huffman + ANS & 1 & 19\\
    \verb|gzip| & LZ77 + Huffman & 1 & 9\\
    \bottomrule
    \end{NiceTabular}
    \caption{Comparison of selected generic compressors.}
    \label{table:generic_compressors}
\end{table*}

\subsection{Datasets} \label{subsec:standard-compression}

We evaluated the compression ratio of \NAME{} on a number of publicly available datasets. Procedurally, we chose datasets with a variety of file formats. We anticipated we would perform well against the chosen data, but for fairness the file formats and datasets were chosen before we created the compressors or tested any competitors. \Cref{table:datasets-summary} summarizes these datasets.

\begin{table}
    \centering
    \begin{NiceTabular}{ l p{5cm} c c }
    \CodeBefore
    \Body
    \toprule
    \multicolumn{1}{l}{Dataset} & \multicolumn{2}{r}{Data Format} & Chunked\\
    \midrule
    \multicolumn{2}{l}{\texttt{binance\_canonical}} & Parquet & No\\
    \multicolumn{2}{l}{\texttt{tlc\_canonical}} & Parquet & No\\
    \multicolumn{2}{l}{\texttt{era5\_flux}} & GRIB & Yes\\
    \multicolumn{2}{l}{\texttt{era5\_precip}} & GRIB & Yes\\
    \multicolumn{2}{l}{\texttt{era5\_pressure}} & GRIB & Yes\\
    \multicolumn{2}{l}{\texttt{era5\_snow}} & GRIB & Yes\\
    \multicolumn{2}{l}{\texttt{era5\_wind}} & GRIB & Yes\\
    \multicolumn{2}{l}{\texttt{psam\_p}} & CSV & No\\
    \multicolumn{2}{l}{\texttt{psam\_h}} & CSV & No\\
    \multicolumn{2}{l}{\texttt{ppmf\_unit}} & CSV & Yes\\
    \multicolumn{2}{l}{\texttt{ppmf\_person}} & CSV & Yes\\
    \bottomrule
    \end{NiceTabular}
    \caption{A summary of the datasets used.}
    \label{table:datasets-summary}
\end{table}

\subsubsection{2020 US Census}
The United States Census Bureau publishes a census every ten years. The unsampled 2020 data is available via the \textit{Privacy-Protected Microdata File} (PPMF)~\cite{ppmf2024}, which anonymizes individual records by adding differentially-private statistical noise. The PPMF files are huge. The household table (\texttt{ppmf\_unit}) is a 54 GB CSV file and the people table (\texttt{ppmf\_person}) is even larger, at 125 GB. We preprocess the files by breaking them into 100 MB chunks, splitting between line breaks. This generates 548 \texttt{ppmf\_unit} files and 1,256 \texttt{ppmf\_person} files.

The Census Bureau also collects the yearly American Community Survey (ACS). The data are available via the \textit{Public Use Microdata Sample} (PUMS)~\cite{pums2023}, a partially-redacted sample of the responses from the ACS. We build our corpus from the 5-year PUMS data from 2023~\cite{pums2023}. There are two sets of CSV files, one for people (10.4 GB) and one for households (4.17 GB). Each is broken down by state. We refer to these datasets as \texttt{psam\_p} and \texttt{psam\_h} in the rest of this section.

\subsubsection{Climate Reanalysis}
One widely-used tool in climate study is the \textit{climate reanalysis}. This is a reconstruction of detailed climate data using existing recorded data and interpolations with modern modelling.

The European Centre for Medium-Range Weather Forecasts (ECMWF) currently maintains the ERA5, the fifth-generation of their global reanalysis dataset~\cite{ERA5}. ERA5 provides hourly estimates of many variables with data from 1940 to the present. For our benchmark, we used 5 datasets from October 1987: 10m u-component of wind (\verb|ERA5_wind|), mean sea level pressure (\verb|ERA5_pressure|), snow density (\verb|ERA5_snow|), downward UV radiation at the surface (\verb|ERA5_flux|), and total precipitation (\verb|ERA5_precip|). For each dataset, there are 720 hourly snapshots, each about 8 MB.

The Hans-Ertel Centre for Weather Research at Bonn University~\cite{HERZ} has published a number of European-centric reanalyses. One such reanalysis is the COSMO-REA6\footnote{Source: Hans-Ertel-Centre for Weather Research.}~\cite{BOLLMEYER15}~\cite{REA6}. We used the total precipitation time series from December 2015 (\verb|REA6_precip|). There are 744 hourly snapshots, each about 5.8 MB.

\subsubsection{NYC Taxi Trip Records}
The New York City Taxi and Limousine Commission (TLC) is the agency responsible
for licensing and regulating New York City's taxis, for-hire vehicles, commuter vans, and paratransit vehicles. The TLC collects and publishes trip record
information for each taxi and for-hire vehicle trip completed by their licensed drivers or vehicles~\cite{TLC}.

These records capture pickup and drop-off dates and times, pickup and drop-off locations, trip distances, itemized fares, rate types, payment types, and driver-reported passenger counts. As of 2025, they also include a congestion fee column.

For our benchmark, we use Yellow and Green Taxi trip data from Q1 2025 (Jan--Mar).  We convert these records to a ``canonical'' Parquet format, with no compression and default encoding. The average size of all records is 228 MB in the canonical format. The size distribution is bimodal with the Yellow records at about 526 MB each and the Green records at about 7.12 MB.

\subsubsection{Binance}
An important resource in quantitative finance is candlestick data, which describe how the price of an asset changes over a given timeframe.

The Binance dataset~\cite{BINANCE} is a collection of 1-minute candlestick data for the top 1000 cryptocurrency trading pairs on \url{binance.com}. For each trading pair, the dataset provides a Parquet file containing fields like \verb|open_time|, \verb|open|, \verb|close|, \verb|high|, and \verb|low|, as retrieved from Binance’s official API endpoint for historical candlestick data.

For our benchmark, we selected 15 Bitcoin candlestick records from the dataset (\verb|BTC-AUD|, \verb|BTC-BIDR|, \verb|BTC-BRL|, \verb|BTC-BUSD|, \verb|BTC-DAI|, \verb|BTC-EUR|, \verb|BTC-GBP|, \verb|BTC-NGN|, \verb|BTC-PAX|, \verb|BTC-RUB|, \verb|BTC-TRY|, \verb|BTC-TUSD|, \verb|BTC-UAH|, \verb|BTC-USDC|, \verb|BTC-USDT|). We convert these records to a ``canonical'' Parquet format, with no compression and default encoding. Each record is about 64 MB.

\subsection{Best Compression Ratio}
\label{subsections:ratio-results}

For each dataset, we generated a test/train split and used the default \NAME{} training script to generate a compressor targeting the best compression without respect to speed. Then we benchmarked the trained compressor on the test set to measure compression ratio, compression speed, and decompression speed.

In addition to the generic compressors, we benchmarked the climate reanalyses against Blosc and Parquet datasets against the default Parquet compression. Note that all the compression workloads, including for \NAME{}, were all single-threaded. The only parallelism used was in training the \NAME{} compressor.

\begin{table*}[h]
    \centering
    \begin{NiceTabular}{l @{ } S[table-format=3.2] @{ } S[table-format=3.2] @{ } S[table-format=3.2] @{ } S[table-format=3.2] @{ } S[table-format=3.2] @{ } S[table-format=3.2] @{ } S[table-format=3.2] @{ } S[table-format=3.2]}
    \CodeBefore
    \rectanglecolor{metabg}{1-7}{13-7}
    \Body
    \toprule
     &
     \multicolumn{1}{@{ }c@{}}{\tt xz-9} &
     \multicolumn{1}{@{ }c@{}}{\tt gzip-6} &
     \multicolumn{1}{@{ }c@{}}{\tt zstd-19} &
     \multicolumn{1}{@{ }c@{}}{Blosc} &
     \multicolumn{1}{@{ }c@{}}{Parquet} &
     \multicolumn{1}{@{ }c@{ }}{\tt OpenZL} \\
     \cmidrule{2-7}
     \tt ppmf-person & 69.10 & 21.20 & 60.15 & & & 116.70\\
     \tt ppmf-unit & 77.05 & 22.11 & 74.99 & & & 102.69\\
     \tt psam-p & 6.75 & 4.61 & 6.47 & & & 8.59\\
     \tt psam-h & 6.27 & 4.43 & 5.98 & & & 7.67\\
     \tt binance-canonical & 2.62 & 1.84 & 2.15 & & 1.25 & 3.01\\
     \tt tlc-canonical & 10.52 & 7.31 & 8.95 & & 8.09 & 11.42\\
     \tt rea6-precip & 21.44 & 14.16 & 18.07 & 16.21 & & 24.82\\
     \tt era5-flux & 8.73 & 5.82 & 7.45 & 7.96 & & 13.62\\
     \tt era5-precip & 11.19 & 7.39 & 9.49 & 7.86 & & 13.00\\
     \tt era5-pressure & 6.58 & 3.75 & 5.60 & 6.01 & & 11.20\\
     \tt era5-snow & 47.76 & 32.16 & 41.01 & 33.97 & & 58.25\\
     \tt era5-wind & 4.88 & 3.12 & 4.12 & 4.17 & & 6.82\\
    \bottomrule
    \end{NiceTabular}
    \caption{Compression ratios on various datasets. Trained \NAME{} compressors are able to beat all tested alternatives on ratio.}
    \label{table:ratio-only}
\end{table*}

\begin{figure*}
    \centering
    \scalebox{.7}{

\pgfplotsset{
    openzlplotstyle/.style={
        mark=openzlplotmark,
        color=openzlplotcolor
    },
    zstdplotstyle/.style={
        mark=zstdplotmark,
        color=zstdplotcolor
    },
    xzplotstyle/.style={
        mark=xzplotmark,
        color=xzplotcolor
    },
    zlibplotstyle/.style={
        mark=zlibplotmark,
        color=zlibplotcolor
    },
    bloscplotstyle/.style={
        mark=bloscplotmark,
        color=bloscplotcolor,
    },
    parquetplotstyle/.style={
        mark=parquetplotmark,
        color=parquetplotcolor
    },
    algo filter/.style n args={1}{
        x filter/.code={%
            \edef\tmpalgo{\thisrow{algorithm}}%
            \edef\tmpalgowanted{#1}%
            \ifx\tmpalgo\tmpalgowanted
            \else
                \def\pgfmathresult{}
            \fi
        },
    },
    speed chart/.style={
        width=6cm,
        height=5cm,
        x unit=\si{\mebi\byte\per\second},
        y unit=\si{\byte\per\byte},
        log ticks with fixed point,
        xmajorgrids,
        ymajorgrids,
        major grid style={ultra thin, densely dotted, black},
        log basis y=2,
        yminorticks=true,
        legend style={
            at={(0.5,0.975)},
            anchor=north,
            cells={
                anchor=west,
            },
        },
        legend columns=2,
        label style={
            font=\footnotesize,
        },
        only marks,
    },
    cspeed chart/.style={
        speed chart,
        xlabel={Compression Speed},
        ylabel={Compression Ratio},
        xshift=-.15cm,
        anchor=east,
    },
    dspeed chart/.style={
        speed chart,
        xlabel={Decompression Speed},
        y unit={},
        yticklabel pos=right,
        xshift=.15cm,
        anchor=west,
    },
    speed chart legend/.style={
    }
}

\begin{tikzpicture}[global]

\coordinate (p1)  at (-5.625,   0);
\coordinate (p2)  at ( 5.625,   0);
\coordinate (p3)  at (-5.625,  -5.5);
\coordinate (p4)  at ( 5.625,  -5.5);
\coordinate (p5)  at (-5.625, -11);
\coordinate (p6)  at ( 5.625, -11);
\coordinate (p7)  at (-5.625, -16.5);
\coordinate (p8)  at ( 5.625, -16.5);
\coordinate (p9)  at (-5.625, -22);
\coordinate (p10) at ( 5.625, -22);
\coordinate (p11) at (-5.625, -27.5);
\coordinate (p12) at ( 5.625, -27.5);

\def\tmpplots{}%

\edef\legendplot{(p2)}%

\pgfplotsforeachungrouped \tablename / \plotpos in {
    {result_tables/perfscatterplot/binance_canonical.csv} /(p1),
    {result_tables/perfscatterplot/era5_flux.csv}         /(p2),
    {result_tables/perfscatterplot/era5_precip.csv}       /(p3),
    {result_tables/perfscatterplot/era5_pressure.csv}     /(p4),
    {result_tables/perfscatterplot/era5_snow.csv}         /(p5),
    {result_tables/perfscatterplot/era5_wind.csv}         /(p6),
    {result_tables/perfscatterplot/ppmf_person.csv}       /(p7),
    {result_tables/perfscatterplot/ppmf_unit.csv}         /(p8),
    {result_tables/perfscatterplot/psam_h.csv}            /(p9),
    {result_tables/perfscatterplot/psam_p.csv}            /(p10),
    {result_tables/perfscatterplot/rea6_precip.csv}       /(p11),
    {result_tables/perfscatterplot/tlc_canonical.csv}     /(p12)
} {

    \eappto\tmpplots{
        \noexpand\begin{semilogxaxis}[
            cspeed chart,
            at=\plotpos
        ]
    }

    \pgfplotsforeachungrouped \algorithm / \algostyle in {
        openzl/openzlplotstyle,
        zstd/zstdplotstyle,
        xz/xzplotstyle,
        zlib/zlibplotstyle,
        blosc/bloscplotstyle,
        parquet/parquetplotstyle
    } {
        \eappto\tmpplots{
            \noexpand\addplot [
                \algostyle,
                algo filter={\algorithm},
            ] table [
                x expr={\noexpand\thisrow{cspeed_mbps}},
                y expr={\noexpand\thisrow{ratio}},
                col sep=comma,
                row sep=newline,
            ] {\tablename};
        }
    }

    \eappto\tmpplots{
        \noexpand\end{semilogxaxis}
    }

    \eappto\tmpplots{
        \noexpand\begin{semilogxaxis}[
            dspeed chart,
            \ifx\plotpos\legendplot%
                speed chart legend,
            \fi%
            at=\plotpos
        ]
    }

    \pgfplotsforeachungrouped \algorithm / \algostyle in {
        openzl/openzlplotstyle,
        zstd/zstdplotstyle,
        xz/xzplotstyle,
        zlib/zlibplotstyle,
        blosc/bloscplotstyle,
        parquet/parquetplotstyle
    } {
        \eappto\tmpplots{
            \noexpand\addplot [
                \algostyle,
                algo filter={\algorithm},
            ] table [
                x expr={\noexpand\thisrow{dspeed_mbps}},
                y expr={\noexpand\thisrow{ratio}},
                col sep=comma,
                row sep=newline,
            ] {\tablename};


            \ifx\plotpos\legendplot%
                \noexpand\addlegendentry{\noexpand\tt\noexpand\footnotesize\algorithm};
            \fi
        }
    }

    \eappto\tmpplots{
        \noexpand\end{semilogxaxis}
    }

}


\tmpplots


\node [txt] at ($(p1)+(0,2.25)$) {Binance (Canonicalized Parquet) Dataset};
\node [txt] at ($(p2)+(0,2.25)$) {ERA5 Flux Dataset};
\node [txt] at ($(p3)+(0,2.25)$) {ERA5 Precip Dataset};
\node [txt] at ($(p4)+(0,2.25)$) {ERA5 Pressure Dataset};
\node [txt] at ($(p5)+(0,2.25)$) {ERA5 Snow Dataset};
\node [txt] at ($(p6)+(0,2.25)$) {ERA5 Wind Dataset};
\node [txt] at ($(p7)+(0,2.25)$) {PPMF Person Dataset};
\node [txt] at ($(p8)+(0,2.25)$) {PPMF Unit Dataset};
\node [txt] at ($(p9)+(0,2.25)$) {PSAM-H Dataset};
\node [txt] at ($(p10)+(0,2.25)$) {PSAM-P Dataset};
\node [txt] at ($(p11)+(0,2.25)$) {REA6 Precip Dataset};
\node [txt] at ($(p12)+(0,2.25)$) {TLC (Canonicalized Parquet) Dataset};

\end{tikzpicture}
}
    \caption{Benchmark results across our test datasets. See also \cref{table:full-results} for a tabular presentation.}
    \label{fig:all-perf-scatterplots}
\end{figure*}
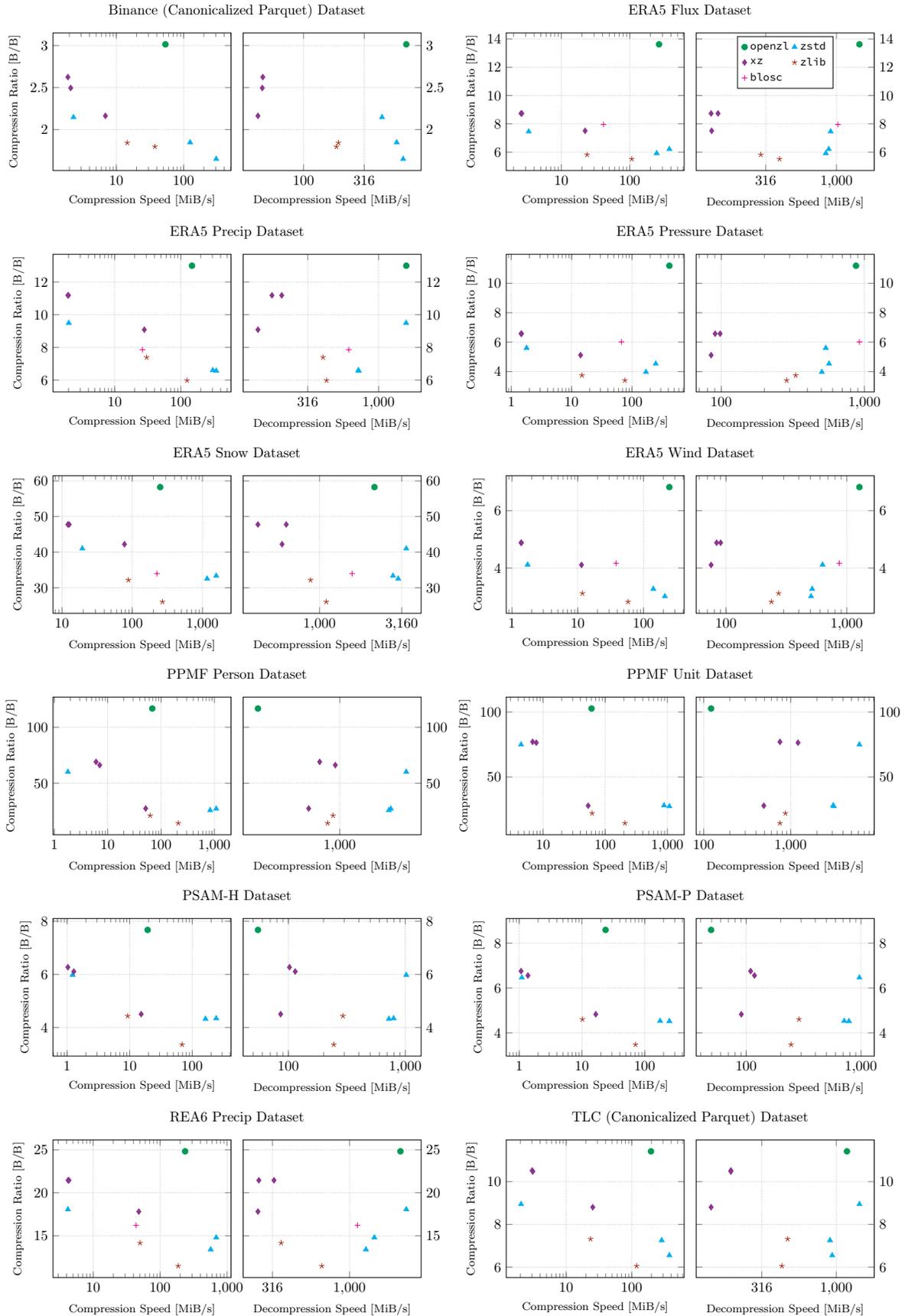

The results we achieved are presented in \cref{fig:all-perf-scatterplots} and \cref{table:ratio-only}. In every case, \NAME{} compressors are able to exceed the best compression ratio offered by \texttt{xz}. Importantly, this is accomplished while compressing at speeds an order of magnitude faster. In the case of \texttt{ppmf\_person}, a trained \NAME{} compressor compresses 55\% better than \texttt{xz-9} and 11 times as fast. Decompression speed is much worse, an artifact of the extra effort spent in parsing. CSV is a particularly time-consuming format to parse; Parquet requires less effort to parse, and thus the speeds are competitive even with \texttt{zstd}. In the case of \texttt{era5\_pressure} and \texttt{era5\_wind}, a trained \NAME{} compressor is even able to beat \verb|zstd| on both compression and decompression speed while achieving ratios double that of \verb|zstd|.

\subsubsection{Trained Compressors}

The training script produces a serialized compressor that is then provided to \NAME{} in addition to the input file when compressing. Note that unlike dictionary-based compression, which uses the dictionary as an out-of-bound communication mechanism between the compressor and decompressor in order to permit a more compact representation of the message, no such smuggling happens here---the compressor does not need to be separately presented to the decoder and the compressed frame remains fully self-describing.

The generated compressors are quite small. \Cref{table:compressor_size} shows the compressor size generated for each test dataset. Most notably, the compressors for the Census datasets are the largest, because we must encode the column clustering instructions in the compressor. Out of these, the \texttt{psam\_p} compressor is the largest, because there are over 200 columns in that table. However, these are outliers. The next largest compressor is 15 KB. 

It's important to note that the numbers shown are for the serialized compressor and \emph{not} the resolved graph in the compressed frame. The compressor is \begin{enuminline} \item much more verbose than the frame format and \item must contain all the possible variants of dynamism it may ever need \end{enuminline}. The actual footprint of the resolved graph is much smaller.

\begin{table}
    \centering
    \begin{NiceTabular}{ l p{4cm} S }
    \CodeBefore
    \Body
    \toprule
    \multicolumn{1}{l}{Dataset} & \multicolumn{2}{r}{Compressor Size}\\
    \midrule
    \multicolumn{2}{l}{\texttt{binance\_canonical}} & 9767\\
    \multicolumn{2}{l}{\texttt{tlc\_canonical}} & 15268\\
    \multicolumn{2}{l}{\texttt{era5\_flux}} & 1226\\
    \multicolumn{2}{l}{\texttt{era5\_precip}} & 1137\\
    \multicolumn{2}{l}{\texttt{era5\_pressure}} & 945 \\
    \multicolumn{2}{l}{\texttt{era5\_snow}} & 890 \\
    \multicolumn{2}{l}{\texttt{era5\_wind}} & 1098\\
    \multicolumn{2}{l}{\texttt{psam\_p}} & 46423\\
    \multicolumn{2}{l}{\texttt{psam\_h}} & 34597\\
    \multicolumn{2}{l}{\texttt{ppmf\_unit}} & 41830 \\
    \multicolumn{2}{l}{\texttt{ppmf\_person}} & 44253 \\
    \bottomrule
    \end{NiceTabular}
    \caption{Trained compressor sizes.}
    \label{table:compressor_size}
\end{table}

\subsubsection{Special Comparisons}
\textbf{Blosc-Btune.}
For the climate reanalysis datasets, we also compared against Blosc (covered in \cref{sections:related_work}). Specifically, we used Btune, their genetic auto-configure add-on, to optimize the ratio for each file. To offer an interesting comparison, we configured the Btune search parameters to prioritize ratio over speed. Reproducibility details are documented in \cref{sections:reproducibility}.

\textbf{PPMF.} Both \verb|PPMF_person| and unit datasets exhibit significant data variation, leading to distinct ``banding'' of a more compressible section and a less compressible section. This is evident, for instance, by running a generic compressor like \texttt{zstd} over the data. We were able to exploit this difference in the data to train different subgraphs to fit each band. We did not do this, but one can imagine a production compressor that dynamically selects between these subgraphs at runtime. A potential implementation could involve auto-detecting the correct band by compressing a 32 KiB block of the input data file using a fast LZ engine. At less than 0.01\% of the size of each file, this would be a negligible addition to compression time and no change for decompression.

\textbf{Parquet.} By default, Parquet compresses its data. For each column, it chooses an ``encoding'' (e.g. tokenize, RLE, etc.) and a backend compressor (\texttt{snappy}, \texttt{gzip}, etc.). We measured the effectiveness of this compression by comparing the file sizes of the default parquet file against the canonicalized parquet, i.e. no encoding, no compression.

\subsection{Compressor Tradeoff Selection}
\label{subsections:pareto-frontier}

\begin{figure*}[ht]
    \centering
    \scalebox{.7}{

\pgfplotsset{
    openzlplotstyle/.style={
        mark=openzlplotmark,
        color=openzlplotcolor
    },
    zstdplotstyle/.style={
        mark=zstdplotmark,
        color=zstdplotcolor
    },
    xzplotstyle/.style={
        mark=xzplotmark,
        color=xzplotcolor
    },
    zlibplotstyle/.style={
        mark=zlibplotmark,
        color=zlibplotcolor
    },
    bloscplotstyle/.style={
        mark=bloscplotmark,
        color=bloscplotcolor
    },
    parquetplotstyle/.style={
        mark=parquetplotmark,
        color=parquetplotcolor
    },
    algo and cpareto filter/.style n args={2}{
        x filter/.code={%
            \edef\tmpalgo{\thisrow{algorithm}}%
            \edef\tmpalgowanted{#1}%
            \edef\tmpcopt{\thisrow{in_algo_cspeed_frontier}}%
            \edef\tmpdopt{\thisrow{in_algo_dspeed_frontier}}%
            \edef\tmpcoptwanted{#2}%
            \ifx\tmpalgo\tmpalgowanted
                \ifx\tmpcopt\tmpcoptwanted
                \else
                    \def\pgfmathresult{}
                \fi
            \else
                \def\pgfmathresult{}
            \fi
        },
    },
    algo and dpareto filter/.style n args={2}{
        x filter/.code={%
            \edef\tmpalgo{\thisrow{algorithm}}%
            \edef\tmpalgowanted{#1}%
            \edef\tmpcopt{\thisrow{in_algo_cspeed_frontier}}%
            \edef\tmpdopt{\thisrow{in_algo_dspeed_frontier}}%
            \edef\tmpdoptwanted{#2}%
            \ifx\tmpalgo\tmpalgowanted
                \ifx\tmpdopt\tmpdoptwanted
                \else
                    \def\pgfmathresult{}
                \fi
            \else
                \def\pgfmathresult{}
            \fi
        },
    },
    speed chart/.style={
        width=6cm,
        height=5cm,
        x unit=\si{\mebi\byte\per\second},
        y unit=\si{\byte\per\byte},
        log ticks with fixed point,
        xmajorgrids,
        ymajorgrids,
        major grid style={ultra thin, densely dotted, black},
        log basis y=2,
        yminorticks=true,
        legend style={
            at={(0.025,0.05)},
            anchor=south west,
            draw=none,
            fill=none,
            cells={
                anchor=west,
            },
        },
        label style={
            font=\footnotesize,
        },
    },
    cspeed chart/.style={
        speed chart,
        xlabel={Compression Speed},
        ylabel={Compression Ratio},
        xshift=-.15cm,
        anchor=east,
    },
    dspeed chart/.style={
        speed chart,
        xlabel={Decompression Speed},
        y unit={},
        yticklabel pos=right,
        xshift=.15cm,
        anchor=west,
    },
    speed chart legend/.style={
    }
}

\begin{tikzpicture}[global]

\coordinate (p1) at (-5.625,   0);
\coordinate (p2) at ( 5.625,   0);
\coordinate (p3) at (-5.625,  -5.5);
\coordinate (p4) at ( 5.625,  -5.5);
\coordinate (p5) at (-5.625, -11);
\coordinate (p6) at ( 5.625, -11);
\coordinate (p7) at ( 0.0  , -16.5);

\def\tmpplots{}%

\edef\legendplot{(p1)}%

\pgfplotsforeachungrouped \tablename / \plotpos in {
    {result_tables/pareto_frontiers/binance_canonical.csv} /(p1),
    {result_tables/pareto_frontiers/era5_precip.csv}       /(p2),
    {result_tables/pareto_frontiers/ppmf_unit.csv}         /(p3),
    {result_tables/pareto_frontiers/tlc_canonical.csv}     /(p4)%
} {

    \eappto\tmpplots{
        \noexpand\begin{semilogxaxis}[
            cspeed chart,
            at=\plotpos
        ]
    }

    \pgfplotsforeachungrouped \algorithm / \algostyle in {
        openzl/openzlplotstyle,
        zstd/zstdplotstyle,
        xz/xzplotstyle,
        zlib/zlibplotstyle
    } {
        \eappto\tmpplots{
            \noexpand\addplot [
                \algostyle,
                mark options={
                    scale=0.75
                },
                algo and cpareto filter={\algorithm}{True},
            ] table [
                x expr={\noexpand\thisrow{compress_speed_mbps}},
                y expr={\noexpand\thisrow{compression_ratio}},
                col sep=comma,
                row sep=newline,
            ] {\tablename};

            \noexpand\addplot [
                \algostyle,
                only marks,
                forget plot,
                mark options={
                    scale=0.75
                },
                algo and cpareto filter={\algorithm}{False},
            ] table [
                x expr={\noexpand\thisrow{compress_speed_mbps}},
                y expr={\noexpand\thisrow{compression_ratio}},
                col sep=comma,
                row sep=newline,
            ] {\tablename};
        }
    }

    \eappto\tmpplots{
        \noexpand\end{semilogxaxis}
    }

    \eappto\tmpplots{
        \noexpand\begin{semilogxaxis}[
            dspeed chart,
            \ifx\plotpos\legendplot%
                speed chart legend,
            \fi%
            at=\plotpos
        ]
    }

    \pgfplotsforeachungrouped \algorithm / \algostyle in {
        openzl/openzlplotstyle,
        zstd/zstdplotstyle,
        xz/xzplotstyle,
        zlib/zlibplotstyle
    } {
        \eappto\tmpplots{
            \noexpand\addplot [
                \algostyle,
                mark options={
                    scale=0.75
                },
                algo and dpareto filter={\algorithm}{True},
            ] table [
                x expr={\noexpand\thisrow{decompress_speed_mbps}},
                y expr={\noexpand\thisrow{compression_ratio}},
                col sep=comma,
                row sep=newline,
            ] {\tablename};

            \noexpand\addplot [
                \algostyle,
                only marks,
                forget plot,
                mark options={
                    scale=0.75
                },
                algo and dpareto filter={\algorithm}{False},
            ] table [
                x expr={\noexpand\thisrow{decompress_speed_mbps}},
                y expr={\noexpand\thisrow{compression_ratio}},
                col sep=comma,
                row sep=newline,
            ] {\tablename};


            \ifx\plotpos\legendplot%
                \noexpand\addlegendentry{\noexpand\tt\noexpand\footnotesize\algorithm};
            \fi
        }
    }

    \eappto\tmpplots{
        \noexpand\end{semilogxaxis}
    }

}

\tmpplots

\node [txt] at ($(p1)+(0,2.25)$) {Binance (Canonicalized Parquet) Dataset};
\node [txt] at ($(p2)+(0,2.25)$) {ERA5 Precip Dataset};
\node [txt] at ($(p3)+(0,2.25)$) {PPMF Unit Dataset};
\node [txt] at ($(p4)+(0,2.25)$) {TLC (Canonicalized Parquet) Dataset};

\end{tikzpicture}
}
    
    \caption{Compression and decompression Pareto frontiers of different algorithms for selected datasets.}
    \label{figure:pareto_frontiers}
\end{figure*}
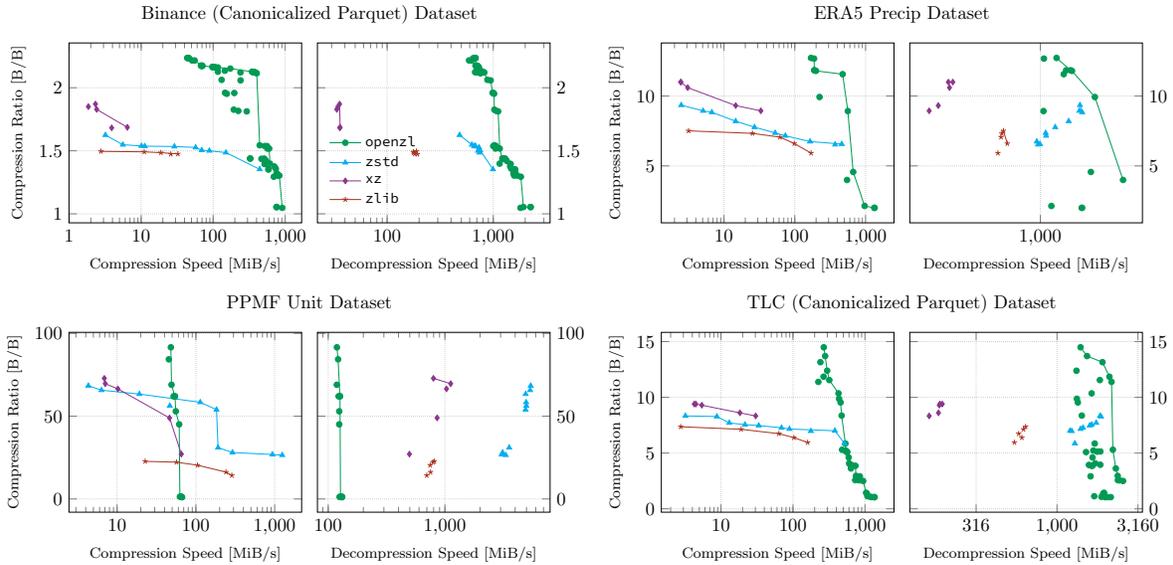

\NAME{} is not limited to pursuing aggressive compression ratios. In addition to the ratio-focused compressors trained in the previous section, we also trained compressors focused on speed, and everything in between. As part of the training process, we generated a Pareto-optimal frontier for some selected datasets, shown in \cref{figure:pareto_frontiers}. Every \NAME{} point on the plot represents a unique compression config.

We plot these against the traditional level system featured by other generic compressors. In many cases, the \NAME{} tradeoff curve for ratio vs.\ compression speed strictly dominates. CSV is a pathological case because \NAME{} needs to spend a lot of energy parsing column information and deserializing numbers. Despite this, some points along the curve are still worthy trade offs. Moreover, the composable nature of the compressor makes it possible to skip the parsing stage in order to reach higher speeds, an option that future iterations of \NAME{} are set to employ. 

\subsection{\texttt{enwik}: A Case Study Where \NAME{} Performs Poorly}
\label{subsections:enwik}

\texttt{enwik} is a dump of English Wikipedia taken on March 6, 2003~\cite{ENWIK}. This is a very important corpus for natural language processing, and there is an ongoing competition to compress the first $10^9$ bytes of it (\texttt{enwik9}) as small as possible~\cite{HUTTER, TEXTCOMPRESSION}.

The file contains an XML dump of English text. This is a domain in which we have not spent time developing good codecs for \NAME{}. In particular, the default behavior for string data is just to run \texttt{zstd}, and we don't expect training to find any gains using the existing suite of numeric and struct-focused codecs.

\begin{table}
    \centering
    \begin{NiceTabular}{ l S S S }
    \CodeBefore
    \Body
    \toprule
    & & \multicolumn{2}{c}{Speeds} \\
    Compressor & {Ratio} & {Comp.} & {Decomp.}\\
    \midrule
    \verb|xz-1| & 2.99 & 10.2 & 71.2\\
    \verb|xz-6| & 3.67 & 1.37 & 84.0\\
    \verb|xz-9| & 3.68 & 1.36 & 79.2\\
    \verb|zstd-1| & 2.44 & 228. & 794.\\
    \verb|zstd-3| & 2.80 & 124. & 649. \\
    \verb|zstd-19| & 3.58 & 1.36 & 589. \\
    \verb|gzip-1| & 2.34 & 54.8 & 199.\\
    \verb|gzip-6| & 2.70 & 14.5 & 199.\\
    \verb|xwrt| & 4.86 & .850 & .811\\
    OpenZL & 3.04 & 45.0 & 449. \\
    \bottomrule
    \end{NiceTabular}
    \caption{Compression results on \texttt{enwik7}. Speeds are measured in Megabytes per second (MB/s).}
    \label{table:enwik7}
\end{table}

To demonstrate this, we benchmarked on the first $10^7$ bytes of \texttt{enwik} (\texttt{enwik7}). As expected, training does not improve the results and the compressor simply uses \texttt{zstd-6} (\cref{table:enwik7}). We add a comparison in the table to \texttt{xwrt}, a compressor specialized to compress natural language~\cite{XWRT}. Unsurprisingly, \texttt{xwrt} achieves the best ratio, beating the next-highest ratio by over 30\% and \NAME{} by almost 60\%.

We present this case study to show that \NAME{} is not a magic bullet for all use cases. The shortcomings are twofold. First, the standard codec library can achieve good results for many datasets, but is not currently optimized for human text. Second, English text is a highly compressible format for which domain-specific transformations are especially useful. \texttt{xwrt} combines multiple such transformations to achieve its results. With this in mind, we are optimistic that porting text-specific codecs and text parsing will improve the performance of \NAME{} compressors in this domain.

\section{\NAME{} at Meta}
\label{sections:openzl_at_meta}

Prior to \NAME{}, Zstandard made up the vast majority of compression use at Meta, since it offered Pareto-optimal performance across a wide variety of use cases and performance tradeoffs. This was the result of a more than decade-long pursuit for compression efficiency at Meta, achieved both by converting uses of compression to Zstd and by optimizing Zstd's performance. As that process went on, it became clear that the headroom for further improvements from Zstd, within Zstd, or even with LZ compression in general, was fundamentally limited. Thus, \NAME{} was born.

\NAME{} has now replaced a meaningful fraction of Zstd use in production at Meta. Much of the data at Meta is serialized using Thrift~\cite{thrift}, both at rest and in transit. A custom parser that understands Thrift, in conjunction with training tools to specialize the compressor for individual use cases, allows \NAME{} to effectively compress a wide range of traffic.

Here is a short survey of these deployments:

\begin{table*}[t]
    \centering
    \begin{NiceTabular}{lcccccc}
    \CodeBefore
    \Body
    \toprule
    Project                   & Use Case          & Data Format                & Trained\\
    \midrule
    Nimble                    & Data warehouse    & Raw Columns                & No\\
    Scribe                    & Data warehouse    & Thrift                     & Yes\\
    Feature storage           & Training data     & Thrift                     & Yes\\
    Log aggregator            & Training data     & Thrift                     & Yes\\
    Embedding storage         & Training data     & Uncompressed \texttt{.zip} & No\\
    PyTorch model checkpoints & Model training    & Float arrays               & No\\
    \bottomrule
    \end{NiceTabular}
    \caption{An overview of major \NAME{} integrations at Meta, as of the time of writing.}
    \label{table:openzl_at_meta}
\end{table*}

\begin{description}
    \item[\href{https://github.com/facebookincubator/nimble}{Nimble}:] Integrating \NAME{} into a columnar database as the backend compressor immediately saved 10\% compressed size compared to Zstd. Most of these gains came from labelling numeric data types as such, and stacks on top of the transformations that Nimble uses to pre-process its data.\footnote{Nimble applies transformations that improve query efficiency by operating on the encoded data, where \NAME{} applies transformations that are useful for compression only.}
    \item[\href{https://engineering.fb.com/2019/10/07/core-infra/scribe/}{Scribe}:] With training, \NAME{} improved compression ratios by $\sim$15\% compared to Zstd. This improves network throughput and improves training data quality through dropping fewer records. Scribe data is also constantly changing. Using the training functionality described in \cref{subsec:training}, we've been able to maintain these ratio wins via regular training runs.
    \item[PyTorch model checkpoints:] \NAME{} leverages type information to save an average of 17\% on storage for model checkpoints, with savings varying based on the floating point data type. Compression with \NAME{} also reduces checkpointing overhead through reduced network traffic, which improves training efficiency.
    \item[Feature storage:] Similar to Scribe, \NAME{} was deployed with training for Thrift data. However, the Feature storage team chose a different trade-off point on the speed-ratio curve. Switching to \NAME{} reduced storage by 10\% and CPU utilization by 5\% compared to Zstd level 6. Moreover, this was done solely by reusing existing components already built for Scribe.
    \item[Log aggregator:] \NAME{} was able to reduce compressed size by 18\% compared to Zstd. In this latency sensitive application, \NAME{}'s performance was improved by sending arrays of integers directly to \NAME{}, cutting out the need to serialize and deserialize.
    \item[Embedding storage:] Compressing bfloat16 embeddings serialized in  PyTorch's \texttt{torch.save()} format reduces compressed size by 30\%. Traditional compressors struggle with floating-point data, so compression was not deemed computationally profitable before this project. The development timeline for this new compressor was on the order of days due to reuse of existing components developed for PyTorch model checkpoints.
\end{description}

Overall, OpenZL has helped Meta bend the curve of AI growth. The compression improvements that OpenZL offers allow Meta to do more with the same amount of hardware---better training data compression means more data can be pushed through the same pipe; smaller data means less compute is spent reading data from the data warehouse; reduced network traffic for model checkpointing improves GPU utilization.

\subsection{Training with Managed Compression}
\label{subsections:training}

As described in \cref{subsections:building_a_compressor_tools}, \NAME{}'s modular treatment of the components of compression, and the development of tools that automate the configuration and composition of those components, mean that \NAME{} lends itself well to offline training. Although the configurations under consideration internally are different and more diverse, treated in the abstract, this training workflow nonetheless has the same overall shape as training a dictionary for Zstandard.

And in fact, Meta's Managed Compression system~\cite{HANDTE18}, which was originally designed to manage dictionaries for Zstandard compression, has proven adept at training compressors for \NAME{}. The trainer accepts a corpus of representative samples and an existing compressor and can configure and parameterize nodes or even construct and replace sub-graphs throughout the compressor. After validation and benchmarking, the resulting compressor can then be re-serialized and deployed to the fleet.

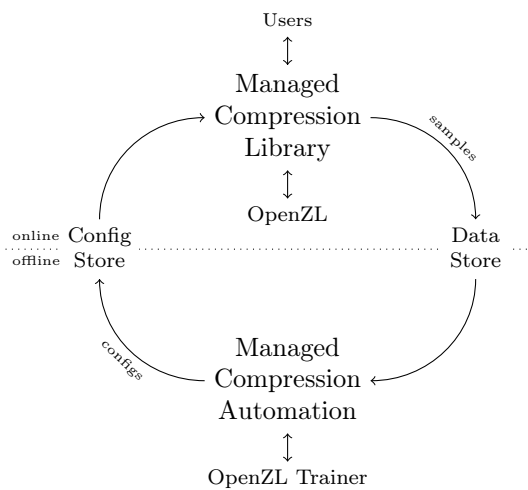
\begin{figure}[h]
    \centering
    \begin{tikzpicture}
        \draw [dotted] (-3.75, 0) -- (3.25, 0);
        \node [txt, tiny, anchor=south west] at (-3.75, 0) {online};
        \node [txt, tiny, anchor=north west] at (-3.75, 0) {offline};
        
        \node [txt] at ( 0,  1.75) (MC_LIB) {Managed\\Compression\\Library};
        
        \node [ss, txt, above=0.375 of MC_LIB.north, anchor=south] (USER) {Users};
        
        \node [fn, txt, below=0.375 of MC_LIB.south, anchor=north] (OZL) {OpenZL};
        
        \node [fn, txt, fill=white] at (2.5, 0) (DATASTORE) {Data\\Store};
        
        \node [txt] at ( 0, -1.75) (TRAINER) {Managed\\Compression\\Automation};
        
        \node [fn, txt, below=0.375 of TRAINER.south, anchor=north] (OZL_TRAINER) {OpenZL Trainer};
        
        \node [fn, txt, fill=white] at (-2.5,  0) (CFGSTORE)  {Config\\Store};
        
        \draw [<->]
              (USER)
              to
              (MC_LIB);
        \draw [<->]
              (MC_LIB)
              to [out=270, in=90]
              (OZL);
        \draw [->]
              (MC_LIB)
              to [out=  0, in= 90]
              node [midway, above, sloped, tiny] {samples}
              (DATASTORE);
        \draw [<->]
              (TRAINER)
              to [out=270, in=90]
              (OZL_TRAINER);
        \draw [->]
              (DATASTORE)
              to [out=270, in=  0]
              (TRAINER);
        \draw [->]
              (TRAINER)
              to [out=180, in=270]
              node [tiny, txt, below, sloped, pos=.5] {configs}
              (CFGSTORE);
        \draw [->]
              (CFGSTORE)
              to [out= 90, in=180]
              (MC_LIB);
    \end{tikzpicture}
    \caption{\NAME{} integrated into Managed Compression.}
\end{figure}

This architecture has proven useful not only in terms of finding useful compression configurations for \NAME{} use cases at Meta, but also in driving broad adoption of \NAME{} at all: in the same way that this systematic approach to training made Zstd dictionaries frictionless for use-cases to adopt, this infrastructure makes \NAME{} easy to integrate, and thousands of trained \NAME{} compressors are deployed to use cases at Meta through Managed Compression.

\section{Conclusions}
\label{sections:conclusions}

This paper describes a graph model of compression, a new conceptual model of compression that encourages composition of small, single-minded codecs. We demonstrate its effectiveness with \NAME{}, a feature-rich implementation of the graph model. The result is a scalable, enterprise-ready system that offers unprecedented performance across diverse datasets. For a wide range of benchmark datasets, we are able to beat the best compression ratio offered by \texttt{xz} at an order of magnitude faster compression, despite needing to parse and understand the data. \NAME{} demonstrates that the compression efficiency unlocked by understanding the data far outweighs the effort spent on parsing, and that this can be achieved via composition of a library of relatively simple codecs.

We hope that the positive results from Meta-internal use cases will motivate data owners to investigate their own wins from using \NAME{}. In particular, we expect the automated training tools presented to be more than adequate to achieve compression wins that justify the resource investment in \NAME{}.

\subsection{Future Work}

The unreasonable effectiveness of our first foray into training leads us to believe that the graph model is uniquely positioned to facilitate ML-guided generation of compressors. We are tempted to view this as ``the next big thing'' in compression. Whereas compression research has up to now eluded those without domain expertise, we believe the future of application-specific compressors will be unlocked via investment in automated learning methods.

We are also emboldened to revisit the space of LZ-based compressors with the graph model. While it's true that LZ on byte streams is essentially a solved problem, the natural extension to typed data still remains an open question. Indeed, our implementation of FieldLZ demonstrates that there is significant progress yet to be made towards a truly generic LZ engine.

\subsection{Contributing to \NAME{}}
\label{sections:conclusions_future_directions}

\NAME{} is open-source. Feedback is welcome and encouraged via Github issues. Many things, including the wire format, are still under development. If you are interested in shaping the evolution of \NAME{}, feel free to contact us!

The codec library is an active area of work. We expect to add more standard codecs and welcome contributors with domain expertise to add domain-specific codecs.

Governance and steering committees are under discussion, as are efforts for hardware acceleration.

\section{Acknowledgements}

We would also like to thank our former interns Timothy Oei, Pedro Valero, Aryan Gandevia, and Faizaan Baig for their contributions to \NAME{}. We would like to thank Dr.\ Evan West for his feedback and suggestions on the manuscript. Finally, we would like to thank Aras Pranckevi\v{c}ius and Takayuki Matsuoka for beta-testing \NAME{}. Their input was useful in shaping the work presented in this paper.

\clearpage
\newpage

\bibliographystyle{assets/plainnat}
\bibliography{paper}

\clearpage
\newpage

\beginappendix

\begin{table*}
    \centering
    \scriptsize
    \begin{NiceTabular}{l c @{ } S @{ } S @{ } S @{ } S @{ } S @{ } S @{ } S @{ } S @{ } S @{ } S @{ } S @{ } }
    \CodeBefore
    \Body
    \toprule
      &  & {\tt{xz-1}} & {\tt{xz-6}} & {\tt{xz-9}} & {\tt{zlib-1}} & {\tt{zlib-6}} & {\tt{zstd-1}} & {\tt{zstd-3}} & {\tt{zstd-19}} & {Blosc} & {Parquet} & {OpenZL}\\
     \cmidrule{2-13}
\multirow{3}{*}{\tt{binance\_canonical}} & R & 2.16 & 2.50 & 2.62 & 1.80 & 1.84 & 1.65 & 1.85 & 2.15 &  & 1.25 & $\bm{3.01}$\\
 & C & 6.88 & 2.10 & 1.91 & 37.4 & 14.6 & $\bm{305.}$ & 124. & 2.31 &  &  & 53.7\\
& D & 42.1 & 45.7 & 46.2 & 187. & 194. & $\bm{663.}$ & 586. & 444. &  &  & 703.\\
\cmidrule{2-13}
 \multirow{3}{*}{\tt{tlc\_canonical}} & R & 8.80 & 10.5 & 10.5 & 6.05 & 7.32 & 6.55 & 7.26 & 8.95 &  & 8.09 & $\bm{11.4}$\\
 & C &  25.4 & 3.08 & 3.03 & 119. & 23.5 & $\bm{375.}$ & 286. & 2.04 &  &  & 197.\\
 & D & 145. & 196. & 196. & 431. & 471. & 938. & 906. & $\bm{1430}$ &  &  & 1180\\
\cmidrule{2-13}
\multirow{3}{*}{\tt{rea6\_precip}} & R & 17.8 & 21.5 & 21.5 & 11.5 & 14.2 & 14.8 & 13.4 & 18.1 & 16.2 &  & $\bm{24.8}$\\
 & C & 48.1 & 4.35 & 4.22 & 187. & 50.2 & $\bm{691.}$ & 572. & 4.19 & 43.8 &  & 237.\\
 & D & 254. & 323. & 258 & 661. & 360. & 1450 & 1280 & $\bm{2340}$ & 1130 &  & 2140\\
\cmidrule{2-13}
\multirow{3}{*}{\tt{era5\_flux}} & R & 7.51 & 8.73 & 8.73 & 5.52 & 5.82 & 6.21 & 5.92 & 7.45 & 7.96 &  & $\bm{13.6}$\\
 & C & 22.2 & 2.63 & 2.57 & 107. & 23.8 & $\bm{379.}$ & 247. & 3.32 & 41.3 &  & 268.\\
 & D & 131. & 146. & 123. & 396. & 292. & 881. & 841. & 909. & 1020 &  & $\bm{1450}$\\
\cmidrule{2-13}
\multirow{3}{*}{\tt{era5\_precip}} & R & 9.09 & 11.2 & 11.2 & 5.99 & 7.39 & 6.57 & 6.61 & 9.50 & 7.86 &  & $\bm{13.0}$\\
 & C & 28.0 & 1.99 & 1.96 & 124. & 30.4 & $\bm{343.}$ & 305. & 2.02 & 26.3 &  & 147.\\
 & D & 141. & 207. & 177. & 429. & 405. & 724. & 720. & 1560 & 616. &  & $\bm{1570}$\\
\cmidrule{2-13}
\multirow{3}{*}{\tt{era5\_pressure}} & R & 5.12 & 6.58 & 6.58 & 3.40 & 3.75 & 4.53 & 3.97 & 5.60 & 6.01 &  & $\bm{11.2}$\\
 & C & 14.1 & 1.48 & 1.45 & 76.3 & 14.8 & 247. & 171. & 1.79 & 66.8 &  & $\bm{417.}$\\
 & D & 84.9 & 98.1 & 90.5 & 286. & 331. & 566. & 504. & 537. & $\bm{923.}$ &  & 873.\\
\cmidrule{2-13}
\multirow{3}{*}{\tt{era5\_snow}} & R & 42.2 & 47.8 & 47.8 & 26.1 & 32.2 & 33.4 & 32.5 & 41.0 & 33.9 &  & $\bm{58.3}$\\
 & C & 77.6 & 12.7 & 12.1 & 269. & 88.2 & $\bm{1570}$ & 1160 & 19.4 & 224. &  & 250.\\
 & D & 592. & 627. & 422. & 1010 & 881. & 2800 & 3010 & $\bm{3370}$ & 1580 &  & 2160\\
\cmidrule{2-13}
\multirow{3}{*}{\tt{era5\_wind}} & R & 4.11 & 4.88 & 4.88 & 2.82 & 3.12 & 3.01 & 3.27 & 4.12 & 4.17 &  & $\bm{6.82}$\\
 & C & 11.4 & 1.39 & 1.37 & 58.3 & 11.8 & 211. & 141. & 1.73 & 38.3 &  & $\bm{248.}$\\
 & D & 75.2 & 90.2 & 83.5 & 237. & 272. & 506. & 517. & 630. & 868. &  & $\bm{1270}$\\
\cmidrule{2-13}
\multirow{3}{*}{\tt{psam\_p}} & R & 4.83 & 6.56 & 6.75 & 3.48 & 4.61 & 4.52 & 4.53 & 6.47 &  &  & $\bm{8.59}$\\
 & C & 16.60 & 1.38 & 1.07 & 71.2 & 10.2 & $\bm{247.}$ & 176. & 1.10 &  &  & 23.8\\
 & D & 90.7 & 118. & 109. & 246. & 288. & 784. & 714. & $\bm{969.}$ &  &  & 49.4\\
\cmidrule{2-13}
\multirow{3}{*}{\tt{psam\_h}} & R & 4.51 & 6.11 & 6.27 & 3.36 & 4.43 & 4.35 & 4.33 & 5.98 &  &  & $\bm{7.67}$\\
 & C & 15.2 & 1.28 & 1.03 & 68.7 & 9.31 & $\bm{241.}$ & 162. & 1.23 &  &  & 19.3\\
 & D & 85.9 & 114. & 102. & 244. & 292. & 794. & 724. & $\bm{1020}$ &  &  & 54.95\\
\cmidrule{2-13}
\multirow{3}{*}{\tt{ppmf\_unit}} & R & 27.9 & 76.4 & 77.1 & 14.5 & 22.1 & 27.5 & 28.2 & 75.0 &  &  & $\bm{103.}$\\
 & C & 53.2 & 7.79 & 6.81 & 207. & 61.6 & $\bm{1070}$ & 882. & 4.43 &  &  & 60.3\\
 & D & 491. & 1210 & 752. & 752. & 869. & 3130 & 3080 & $\bm{6170}$ &  &  & 121.\\
\cmidrule{2-13}
\multirow{3}{*}{\tt{ppmf\_person}} & R & 27.5 & 66.2 & 69.1 & 14.4 & 21.2 & 27.3 & 25.9 & 60.2 &  &  & $\bm{117.}$\\
 & C & 51.4 & 7.09 & 6.06 & 209. & 62.5 & $\bm{1080}$ & 829. & 1.80 &  &  & 68.4\\
 & D & 483. & 902. & 626. & 754. & 855. & 3310 & 3160 & $\bm{4750}$ &  &  & 147.\\
    \bottomrule
    \end{NiceTabular}
    \caption{Compression benchmark results on our test datasets. \textbf{R} represents compression ratio, measured as $\nicefrac{\mathrm{size}_\mathrm{orig}}{\mathrm{size}_\mathrm{comp}}$. \textbf{C} is compression speed measured in MiB/s. \textbf{D} is decompression speed, measured similarly. \textbf{Bold} cells represent the best performance for each evaluation axis. See also 
    \cref{fig:all-perf-scatterplots} for a graphical presentation.}
    \label{table:full-results}
\end{table*}

\section{Standard Component Library}
\label{sections:components}

Standard components in OpenZL are components that are pre-registered into the compressor. These are well tested and reusable components frequently used in compressors. There are both standard codecs (standalone nodes that can be composed) and standard graphs (prebuilt compression graphs).

The authoritative documentation for these components can be found in the \NAME{} \href{https://facebook.github.io/openzl/}{documentation website}, but they are briefly described in the following sections. As of the time of writing, \NAME{} is still in development, so this list is expected to evolve over time. Indeed, \NAME{}'s format versioning scheme allows safe additions and removals over time.

\subsection{Standard Codecs}

Standard Codecs in OpenZL can be classified into 6 broad categories: conversion, data restructuring, LZ, representation interpreting and transforming.

\begin{description}
\item[Conversion:]
The \NAME{} implementation approximates \textit{Message Sets} with a simple type system that allows \textit{serial} streams of bytes, fixed-width \textit{struct} streams, \textit{numeric} streams, and streams of variable length \textit{strings}. The conversion codecs handle converting from one type to another. The following conversion codecs exist:

\begin{itemize}
\item Conversion from \textit{serial} to \{8, 16, 32, 64\}-bit \textit{numeric} in both big-endian and little-endian formats.
\item Conversion from \textit{numeric} to \textit{serial} in the little-endian format.
\item Conversion from \textit{serial} to fixed-size \textit{structs} of any width.
\item Conversion from \textit{struct} to \textit{serial}.
\item Conversion from \textit{serial} to \textit{string} by adding a lengths stream that describes the length of each string in the content stream.
\item Conversion from a \textit{string} stream to a \textit{serial} and \textit{numeric} stream describing the content and lengths respectively.
\end{itemize}

\item[Data Restructuring:]
Data restructuring codecs in OpenZL perform the role of extracting structure out of the original data. These codecs can be used to separate out homogenous data, and then piece together correlated data so that entropy stages of the compression can better exploit the structure of the data and achieve better compression ratios. The logic that drives these codecs often understands the input data schema, but because the decisions are written into the compressed frame, the decoder can be oblivious to the schema. In this category, OpenZL has the following codecs:

\begin{itemize}
\item Dispatch
\item Dispatch string
\item Concat
\item Interleave
\item Split
\end{itemize}

\item[LZ:]
These codecs implement LZ compression. Today, Field LZ is the only native LZ codec, which runs LZ on \textit{struct} streams and finds matches that match runs of entire \textit{structs} rather than bytes. OpenZL has the following copy-based codecs:

\begin{itemize}
\item Field LZ
\item Zstd
\end{itemize}

\item[Representation Interpreting:]
Representation interpreting codecs perform the role of converting the data from a known representation, to the original data representation. When data is already pre-compressed, or transformed to a representation other codecs are not designed to work with, it is often necessary to restore the original data representation to achieve better compression. OpenZL has the following codecs:

\begin{itemize}
\item Bitunpack
\item Parse Int
\end{itemize}

\item[Transform:]
Transform codecs transform the data (without compressing it) so that following codecs can better exploit the data. Certain patterns, such as strictly increasing data can only be captured using such codecs.

\begin{itemize}
\item Delta
\item Divide By
\item Float Deconstruct
\item Prefix Encoding
\item Range Pack
\item Transpose
\item Tokenize
\item Zigzag
\end{itemize}

\item[Entropy:]
Entropy codecs exploit similarity inside an input to compress the data. Entropy codecs are the backbone of compression and in the majority of cases, and will be used as the final stage in a compressor. \NAME{} has the following codecs:

\begin{itemize}
\item Constant
\item Huffman
\item FSE
\item Bitpack
\end{itemize}

\end{description}

\subsection{Standard Graphs}

Standard graphs are graphs that are automatically registered in the compressor with canonical names. These are common graphs used in the design of a compressor. 

The most fundamental graph is the store graph. This graph directly writes the result to the compressed output and does no transformation on the data. There is a store graph variant for each data type of OpenZL. 

\NAME{} provides these builtin standard graphs:

\begin{itemize}
\item Bitpack
\item Compress
\item Entropy
\item FieldLZ
\item FSE
\item Huffman
\item Zstd
\end{itemize}

Additionally, OpenZL has some specialized graphs to support advanced use cases.

\begin{description}
\item[Generic Clustering:] this graph is a multi-input graph that interprets a config provided and clusters them according to that config then sends them to the successors defined in this config. This graph can be used with the training api provided.
\item[SDDL:] this graph takes a description written in the Simple Data Description Language and applies it to its input, eventually producing a set of dispatch instructions which it sends along with the input to a dispatch codec.
\end{description}

\section{Reproducibility}
\label{sections:reproducibility}

\NAME{} is open-source and available \href{https://www.github.com/facebook/openzl}{here}. This includes the library, training executables, and other tools.

For \cref{subsections:ratio-results}, the benchmark script is included at \texttt{openzl/\allowbreak{}contrib/\allowbreak{}reproducibility/\allowbreak{}watermark} for completeness. The expected time for completion is around 12-16 hours. Most of the time is spent in training, so ensure you have a strong machine with plenty of memory and many cores for the best results. The \texttt{lzbench} script we used is available at \texttt{openzl/\allowbreak{}contrib/\allowbreak{}reproducibility/\allowbreak{}lzbench}. This script will take around 5 days to complete the entire benchmark.

Blosc-specific benchmarks can be found \href{https://www.github.com/Victor-C-Zhang/blosc2-bench}{here}.

For \cref{subsections:pareto-frontier}, the results for the Pareto-frontier figures were generated by \texttt{openzl/\allowbreak{}contrib/\allowbreak{}reproducibility/\allowbreak{}figures/\allowbreak{}make\allowbreak{}-\allowbreak{}pareto\allowbreak{}-\allowbreak{}optimal\allowbreak{}-\allowbreak{}figures\allowbreak{}.py}. The invocations used to generate the charts are saved \href{https://github.com/facebook/openzl/releases/download/openzl-sample-artifacts/2025-09-22-figures-artifact.tar.gz}{here}.

The datasets for \cref{subsections:ratio-results} and \cref{subsections:pareto-frontier} are publicly available. \texttt{openzl/\allowbreak{}contrib/\allowbreak{}reproducibility/\allowbreak{}dataset\allowbreak{}\_\allowbreak{}manager} details this process and provides some convenience scripts to download and process the data. Some datasets require you to apply for an API key before downloading.

For \cref{subsections:enwik}, the dataset is publicly available \href{https://mattmahoney.net/dc/textdata.html}{here}. For speed, we tested with the first $10^7$ bytes. We used the same \texttt{lzbench} settings as \cref{subsections:ratio-results}. The \texttt{xwrt} results were gathered by downloading and building the binary and running with command options
\begin{verbatim}
xwrt -l14 -b255 -m96 -s -e40000 -f200
\end{verbatim}

Instructions to disable boost varies by CPU. On our machine it is as follows:
{\small%
\begin{verbatim}
echo "passive" | sudo tee \
  /sys/devices/system/cpu/amd_pstate/status
echo 0 | sudo tee \
  /sys/devices/system/cpu/cpufreq/boost
\end{verbatim}
}

\end{document}